\documentclass[11pt,twoside]{article}   
\usepackage{epsfig}
\date{}

\title{The Magnetospheric Eternally Collapsing
Object (MECO) Model of Galactic Black Hole Candidates and Active
Galactic Nuclei}  

\author{Stanley L. Robertson\footnote{Dept. of Physics, Southwestern
Oklahoma State University, Weatherford, OK 73096} and Darryl J.
Leiter\footnote{MARC, P.O. 7466, Charlottesville, VA 22901}}

\begin{document}           

\maketitle                 

\begin{abstract}

The spectral, timing, and jet formation properties of  neutron
stars in low mass x-ray binary systems are influenced by the
presence of central magnetic moments. Similar features shown by
the galactic black hole candidates (GBHC) strongly suggest that
their compact cores might be intrinsically magnetic as well.  We
show that the existence of intrinsically magnetic GBHC is
consistent with a new class of solutions of the Einstein field
equations of General Relativity. These solutions are based on a
strict adherence to the Strong Principle of Equivalence (SPOE)
requirement that the world lines of physical matter must remain
timelike in all regions of spacetime. The new solutions emerge
when the structure and radiation transfer properties of the energy
momentum tensor on the right hand side of the Einstein field
equations are appropriately chosen to dynamically enforce this
SPOE requirement of timelike world line completeness. In this
context, we find that the Einstein field equations allow the
existence of highly red shifted, Magnetospheric, Eternally
Collapsing Objects (MECO).  MECO necessarily possess intrinsic
magnetic moments and they do not have trapped surfaces that lead
to event horizons and curvature singularities. Their most striking
features are equipartition magnetic fields, pair plasma
atmospheres and extreme gravitational redshifts. Since MECO
lifetimes are orders of magnitude greater than a Hubble time, they
provide an elegant and unified framework for understanding a broad
range of observations of GBHC and active galactic nuclei. We
examine their spectral, timing and jet formation properties and
discuss characteristics that might lead to their confirmation.
\end{abstract}

\section{Introduction}
The evidence for the existence of massive objects that are compact
enough to be black holes is strong, although there is as yet no
direct evidence of any mass that is contained within its
Schwarzschild radius. Supermassive compact objects have been found
in the nuclei of most galaxies, while objects of stellar mass are
abundant within our own and other galaxies. They are commonly
called black holes nowadays, though no compelling evidence of an
event horizon, the quintessential feature of a black hole, has yet
been found. It is true that in quiescent states of galactic black
hole candidates (GBHC) there are no thermal soft x-ray emission
peaks, such as those possibly seen in some neutron star (NS)
systems. These less luminous x-ray emissions of quiescent GBHC
compared to quiescent NS have been attributed to the presence of
an event horizon [Narayan, Garcia \& McClintock 1997], but not
compellingly so [Abramowicz, Kluzniak \& Lasota 2002]. Although no
thermal peak has been discerned for quiescent GBHC, both GBHC and
NS systems produce power-law soft x-ray spectra of photon index
$\sim 1.7$ or softer for lower luminosities. We show that these
emissions from quiescent NS systems are clearly magnetospheric in
origin. In the magnetic, eternally collapsing object (MECO) model
of BHC which we explore here, the similar quiescent spectra of
GBHC originate in the same way, but at lower luminosity due,
primarily, to slower rates of spin for the GBHC. For a given
magnetic moment, the radiation rate depends on the fourth power of
the spin frequency. The MECO model is fully compatible with
General Relativity.\footnote{.. and perhaps other gravitational
theories; the prime requirement being that objects with extremely
large gravitational redshifts be encompassed by the theory.} The
great strength of the MECO model is that it allows a unified
description of all of the various spectral, luminosity, and rapid
variability states of x-ray novae, whether NS or GBHC. We show
that it can be extended also to the realm of active galactic
nuclei (AGN). In following sections and appendixes we provide the
general relativistic theoretical basis for the existence of MECO,
describe their physical characteristics, their interactions with
accretion disks, their abilities to power jets and describe how
the various spectral states are related to the MECO -
magnetosphere - disk interactions. To avoid interrupting the
presentation of ideas that are closely tied to observational
issues, we have placed several separate topics in the appendixes
and have referred to them as needed. The details in Appendix D are
central to our understanding of the physics that dictates the
large redshifts of the MECO model. The radiating MECO model is
necessarily described by the Vaidya metric, for which there is no
transformation to the Kerr-Schild coordinates used in many black
hole models.
%
%
%
%
%
%
%
%
%
%
%
\section{The Case for Intrinsic, Central Magnetic Moments:}
The similarities of NS and GBHC properties, particularly in low
and quiescent states, have been previously noted, [e.g. van der
Klis 1994, Tanaka \& Shibazaki 1996]. Jets and their synchrotron
emissions in NS, GBHC and AGN also have obvious magnetic
signatures. It is axiomatic that astrophysical objects of stellar
mass and beyond have magnetic moments if they are not black holes,
but an intrinsic magnetic moment is not a permissible attribute of
a black hole. Yet in earlier work, [Robertson \& Leiter 2002] we
presented evidence for the existence of intrinsic magnetic moments
of $\sim 10^{29-30}$ gauss cm$^3$ in the GBHC of LMXB. These
findings are recapitulated and extended in Table 1 and Appendix C.
Calculated values in Table 1 (see also Section 8) have been
corrected using a more recent correlation of spin-down energy loss
rate and soft x-ray luminosity [Possenti et al. 2002], but results
are scarcely changed from the previous work [Robertson \& Leiter
2002] except for new additions listed in bold font. Observational
luminosity and spectral data were analyzed to obtain magnetic
moments and spins of the objects in the table. These magnetic
moments and spin rates were then used to predict quiescent
luminosities $\sim 10^{3-6}$ times less luminous than those
analyzed. The accurate predictions of quiescent luminosities and
predicted spin rates comparable to those observed in NS burst
oscillations [Strohmayer \& Markwart 2002, Chakrabarty et al.
2003] are very powerful confirmation of the magnetospheric origin
of quiescent power-law luminosity. The magnetic moments are
reassuringly similar in magnitude to those determined from the
spin-down rates of similarly rotating millisecond pulsars.
Combined with rotation rates in the range 1 - 40 Hz, the GBHC
magnetic moments provide a robust unified mechanism for the X-ray
spectral state switches observed in GBHC and NS, a common origin
of quiescent power-law emissions as spin-down luminosity, and a
unified driving mechanism for the ubiquitous low-state jets and
synchrotron emissions of both. We have shown [Robertson \& Leiter
2004 and Section 10] that the jet mechanism scales up to the AGN
without difficulty. Magnetosphere topology also serves to
stabilize the inner accretion disk in LMXB [Arons et al. 1984].

\begin{table*}
\footnotesize
\begin{center}
\caption{$^a$Calculated and Observed Quiescent Luminosities}
\begin{tabular}{lrrrrrrrr} \hline
Object & m & $L_{min}$ & $L_c$ & $\mu_{27}$ & $\nu_{obs}$ & $\nu_{calc}$ & log (L$_q$) & log (L$_q$) \\
    & M$_\odot$ & $10^{36}$erg/s & $10^{36}$erg/s & Gauss cm$^3$    & Hz      &  Hz~ & erg/s & erg/s \\
    & & & & & obs. & calc. & obs. & calc. \\ \hline
\bf {NS} \\
Aql X-1 & 1.4 & 1.2 & 0.4 & 0.47 & 549 & 658 & 32.6  & 32.5  \\
4U 1608-52& 1.4 & 10 &2.9 & 1.0 & 619 & 534 & 33.3 & 33.4 \\
Sax J1808.4-3658& 1.4 & $^b$0.8 & 0.2 & 0.53 & 401 & 426 & 31.8-32.2 & 32 \\
Cen X-4 & 1.4 & 4.4 & 1.1 & 1.1 & & 430 & 32.4 & 32.8 \\
KS 1731-26 & 1.4 & & 1.8 & 1.0 & 524 & & $^c$\bf{32.8} & 33.1  \\ 
\bf{XTE J1751-305} & 1.4 & & 3.5 & 1.9 & 435 & & $<$34.3 & 33.7 \\
\bf{XTE J0929-314} & 1.4 & & 4.9 & 8.5 & 185 & & & 33.1 \\
4U 1916-053 & 1.4 & $\sim$14 & 3.2 & 3.7 & 270 & 370 & & 33.0 \\
\bf{4U1705-44} & 1.4 & 26 & 7 & 2.5 & & 470 & & 33.7 \\
4U 1730-335 & 1.4 & 10 & & 2.5 & 307 & & & 32.9 \\
\bf{GRO J1744-28} & 1.4 & & 18 & 13000 & 2.14 & & & 31.5 \\
Cir X-1 & 1.4 & 300 & 14 & 170 & & 35 &  & 32.8 \\ \hline
\bf{GBHC} \\
GRS 1124-68 & 5 & 240 & 6.6 & 720 & & 16 & $<32.4$ & 32.7 \\
GS 2023+338 &7 & 1000 & 48 & 470 & & 46 & 33.7 & 34 \\
XTE J1550-564 & 7 & $^d$90 & 4.1 & 150 & & 45 & 32.8 & 32.2 \\
GS 2000+25 & 7 & & 0.15 & 160 & & 14 & 30.4 & 30.5\\
GRO J1655-40 & 7 & 31 & 1.0 & 250 & & 19 & 31.3 & 31.7 \\
A0620-00 & 4.9 & 4.5 & 0.14 & 50 &  & 26 & 30.5 & 30.2 \\
Cygnus X-1 & 10 & & 30 & 1260 & & 23 &  & 33 \\
GRS 1915+105 & 7 & & 12 & 273 & $^e$27 & & & 33 \\
\bf{XTE J1118+480} & 7 & & 1.2 & 1000 & & 8 & & 31.5 \\
\bf{LMC X-3} & 7 & 600 & 7 & 860 & & 16 & & 33 \\ \hline
\end{tabular}
\end{center}
$^a$New table entries in bold font are described in Appendix C. \\
Equations used for calculations of spins, magnetic moments and $L_q$ are in Section 8\\
Other tabular entries and supporting data are in Robertson \& Leiter [2002]\\
$^b$2.5 kpc, $^c$ [Burderi et al. 2002], $^d$d = 4 kpc \\
$^e$GRS 1915+105 Q $\approx 20$ QPO was stable
for six months and a factor of five luminosity change. \hfill \\
\end{table*}

There is a plethora of piece meal models of the various spectral
and timing characteristics of LMXB. For example, comptonizing
coronae near event horizons, bulk flow comptonization and magnetic
flares on accretion disks have all been invoked to explain the
hard spectral tail of low state GBHC. But the observed
ingress/egress times for dipping sources imply large radiating
regions [Church 2001] that are inconsistent with the compact
corona models and can be consistent with bulk comptonization
models only for large scale outflows. Similarly, radiatively
inefficient advective accretion flows (ADAF) at high accretion
rates have been proposed to explain the quiescent power-law
emissions of GBHC, [Narayan et al. 1997, Garcia et al. 2001],
while ignoring the fact that we have explained the similar
emissions of accreting millisecond pulsars and the NS of LMXB via
magnetospheric spin-down. It is also very disconcerting that ADAF
models are clearly excluded for NS systems. The ADAF accretion
rates are so high that the only way to accommodate them would be
to have a very efficient magnetic propeller mechanism to prevent
the flow from reaching the NS surface and producing very
non-quiescent levels ($10^{35-36}~erg/s$) of emission. But if a
magnetic propeller expels a low-state ADAF flow, then quiescent NS
would necessarily be extremely strong radio sources, contrary to
observations.

Stating the similarities of GBHC and NS systems more bluntly, Cir
X-1, a burster and a `Z track' source with a magnetic moment
similar to those we have found for GBHC [Table 1 and Iaria et al.
2001], exhibits all of the x-ray spectral and timing
characteristics that have been proposed at various times as
distinguishing features of black holes. Both Cir X-1 and Cygnus
X-3, a pulsar microquasar, [Brazier et al. 1990, Mitra 1998] are
strong jet and radio sources similar to GBHC Cygnus X-1 and GX
339-4. A unified model of GBHC and NS is clearly needed. It is
difficult to understand how such common behaviors of obviously
magnetic origin could be produced
both with and without event horizons.

Others have reported evidence for strong magnetic fields in GBHC.
A field in excess of $10^8$ G has been found at the base of the
jets of GRS 1915+105 [Gliozzi, Bodo \& Ghisellini 1999, Vadawale,
Rao \& Chakrabarti 2001]. A recent study of optical polarization
of Cygnus X-1 in its low state [Gnedin et al. 2003] has found a
slow GBHC spin and a magnetic field of $\sim 10^8$ gauss at the
location of its optical emission. These field strengths exceed
disk plasma equipartition levels, but given the $r^{-3}$
dependence of field strength on magnetic moment, the implied
magnetic moments are in very good agreement with those we report
in Table 1. A recent correlation [Mauche et al. 2002, Warner \&
Woudt 2003] of quasi-periodic oscillation (QPO) frequencies
extending over six orders of magnitude in frequency, from dwarf
novae to neutron stars shows points for GBHC squarely in the
middle of the correlation line. If the higher of the correlated
frequencies is generated where the inner radius of an accretion
disk interacts with a magnetosphere [Goodson, Bohm \& Winglee
1999, Titarchuk \& Wood 2002], this would be additional evidence
of intrinsic magnetic moments for GBHC. A relativistic
frame-dragging explanation of
these QPO is surely not applicable.

Although there are widely studied models for generating magnetic
fields in accretion disks, they can produce equipartition fields
at best [Livio, Ogilvie \& Pringle 1999], and perhaps at the
expense of being too luminous [Bisnovatyi-Kogan \& Lovelace 2000]
in quiescence and in any case, too weak and comoving in accretion
disks to drive jets. While tangled magnetic fields in accretion
disks are very likely responsible for their large viscosity, [e.g.
Hawley, Balbus \& Winters 1999] the highly variable mass accretion
rates in LMXB make it unlikely that disk dynamos could produce the
stability of fields needed to account for either spectral state
switches or quiescent spin-down luminosities. Both require
magnetic fields co-rotating with the central object. Further, if
disk dynamos produced the much larger apparent magnetic moments of
GBHC, they should produce them also for the NS systems and cause
profound qualitative spectral and timing differences from GBHC due
to interactions with the intrinsic NS magnetic moments. Such
qualitative differences as have been observed, e.g., the hard
spectral tail of the steep power law (intermediate) state, lack of
surface bursts for GBHC and stronger GBHC jets, (excepting NS Cir
X-1 and Cygnus X-3) are easily explained by differences in masses,
magnetic field strengths and surface redshift. Not only are there
are no observed differences that require explanation in terms of
event horizons [Abramowicz, Kluzniak \& Lasota 2002], there appear
to be none that would be consistent with having two different
magnetic structures for NS and only one GBHC.

It has been suggested that stable magnetic fields could be
produced by electrically charged, rotating black holes [Punsly
1998, Gnedin et al. 2003], however the charge necessary to endow
Cygnus X-1 with a $10^8$ G magnetic field, well out in the
accretion disk, was found to be $5 \times 10^{28}$ esu [Gnedin et
al. 2003]. Due to the large charge/mass ratios of accreting
protons or electrons, this quantity of charge on a black hole
would produce electric forces at least $\sim 10^6$ larger than the
gravitational attraction of a $10 M_\odot$ GBHC, thus causing
charges of one sign to be swallowed and the other to be blown
away. At accretion rates needed to account for the x-ray
luminosity of Cygnus X-1, the original charge would be neutralized
in a fraction of one second. Thus it appears that current black
hole models are unable to offer unified explanations of such
obviously magnetic phenomena as jets, spectral state switches and
quiescent synchrotron emissions and if they could, it seems
unusually generous for nature to have provided different
mechanisms by which
NS and GBHC could produce such strikingly similar phenomena.

In Table 1, we have provided solid evidence for the applicability
of a spinning magnetic moment model of LMXB and we have presented
reasons for considering these magnetic moments to be intrinsic to
the central object rather than being generated within an accretion
disks. While accommodating intrinsic magnetic moments in models of
GBHC will require abandoning the currently popular black hole
theory of GBHC, it will also greatly simplify the problem of
understanding the spectral, timing and jet ejection mechanisms of
compact objects. As noted by Abramowicz, Kluzniak \& Lasota
[2002], it is unlikely that we will ever find direct observational
proof of an event horizon, however, we may be able to definitively
determine whether or not GBHC have intrinsic magnetic moments. We
regard Table 1 as strongly suggestive, if not yet definitive. If
GBHC are not black holes, they would almost certainly be
magnetized, and likely to at least a degree similar
to their compact NS cousins.

In our MECO model, we have found that it is possible to virtually
stop and maintain a slow, (many Hubble times!) steady collapse of
a compact physical plasma object outside of its Schwarzschild
radius with photon pressure generated by synchrotron radiation
from an equipartition surface magnetic field. To control the rate
of collapse, the object must radiate at the local Eddington limit,
but from a highly redshifted surface. (see Appendices D and E.)
There is recent evidence for the presence of such extreme magnetic
fields in gravitational collapse. Equipartition magnetic fields
have been implicated as the driver of GRB 021206 [Coburn \& Boggs
2003] and fields much in excess of those expected from mere flux
compression during stellar collapse have been found in magnetars
[Ibrahim, Swank \& Parke 2003]. Kluzniak and Ruderman [1998] have
described the generation of $\sim 10^{17}$ G magnetic fields via
differential rotation in neutron stars. In Appendix D, we show
that surface drift currents within a pair plasma at the MECO
surface generate its required fields. Drift currents proportional
to ${\bf g\times B}/B^2$ occur for plasmas
at rest in gravitational and magnetic fields.

The equatorial poloidal magnetic field needed for a stable rate of
collapse of the exterior surface is $\sim 10^{20}~gauss$. Fields
of this magnitude are strong enough to create bound
electron-positron pairs out of the quantum electrodynamic vacuum.
This assures sufficient photon pressure from annihilation
radiation to stabilize the collapse rate. The magnetic field of
the interior is approximately what one would expect from flux
compression during collapse, $\sim 2.5\times
10^{13}\sqrt{7M_\odot/M}$ gauss and its radial component is
continuous across the surface boundary. The poloidal field is
discontinuous across the surface and much stronger externally due
to the surface drift currents. As shown in Appendix D, at the MECO
surface radius $2R_g=2GM/c^2$, the ratio of poloidal field on the
surface to the poloidal field just under the MECO surface is given
by
\begin{equation}
B_{\theta,S^+}/B_{\theta,S^-}=(1+z_s)/(2ln(1+z_s))=10^{20}/(2.5\times
10^{13})\sqrt{7M_\odot/M})
\end{equation}
where $z_s$ is the surface redshift. This has the solution
\begin{equation}
1+z_s = 1.5\times 10^8\sqrt{M/7M_\odot}
\end{equation}
The distantly observed field is reduced by a surface redshift of
by a factor of $3(1+z_s) =4.5\times 10^8\sqrt{M/7M_\odot}$ to a
level which agrees well with the observed magnetic moments shown
in Table 1. The surface luminosity is reduced below the
conventional Newtonian Eddington limit by $(1+z_s)$ when distantly
observed, and the decay lifetime is extended by the same factor.

\section{The Strong Principle of Equivalence}
Astrophysicists nowadays generally accept the inevitability of the
curvature singularities of black holes \footnote {The modern
notion of a black hole began with Hilbert's error of application
of boundary conditions for the solutions of Einstein's field
equations of General Relativity for a mass point. Hilbert's
solution has been erroneously attributed to Schwarzschild, however
Schwarzschild's preceding original solution had no event horizon.
[see Abrams 1979, 1989] Nevertheless, Hilbert's solution is
analytically extendible through the horizon to a central
singularity, hence the modern black hole. The fact that the radius
of the event horizon, which is directly proportional to the
gravitational mass, can be changed arbitrarily by (generally
singular) coordinate transformation strongly suggests that the
horizon is unphysical.}, however, if the GBHC are confirmed as
intrinsically magnetized, this will be nature's way of telling us
that such singularities are not really permitted to exist. For
black holes to exist, gravity must be able to do what no other
force of nature can do; namely, to accelerate the physical
three-speed of a finite mass to exactly the speed of light. But
this means that horizon crossing geodesics, in realistic
coordinates, would become null rather than timelike. In General
Relativity (GR) the Strong Principle of Equivalence (SPOE)
requires that Special Relativity (SR) must hold locally for all
time-like observers in all of spacetime. This SPOE requirement is
a tensor relationship that implies that (i) the spacetime manifold
for observers located in field-free regions, distant from
gravitating masses, must approach the flat spacetime of SR
\footnote{This eliminates Kruskal-Szekeres coordinates from the
real world of astrophysics.} and (ii) the spacetime world lines of
massive matter must always be timelike. \footnote{Models of
gravitational collapse that lead to the development of event
horizons and central curvature singularities inevitably abandon
the SPOE requirement for timelike world line completeness. The
vanishing of the metric time coefficient, $g_{tt}$, at the
Schwarzschild radius is sufficient to cause the timelike physical
three-speed of particles in radial free fall to approach the speed
of light there. In obvious notation, [Landau \& Lifshitz 1975].
$V^2 = (\frac{dl}{d\tau_s})^2 = c^2\frac{({g_{tr}g_{tr}} -
g_{rr}g_{tt})v^r v^r}
        {(g_{tt} + g_{tr} v^r)^2}$ where $v^r=\frac{dr}{d\tau}$.
In non-singular Finkelstein or Kerr-Schild coordinates, for which
$g_{tr} \neq 0$, we find $V \rightarrow c$ as $g_{tt} \rightarrow
0$.  It has been shown [Leiter \& Robertson 2003] that $ds^2
\rightarrow 0$ at surfaces of infinite redshift. In
Kruskal-Szekeres coordinates, in which $g_{tt}$ does not vanish,
there is no surface of infinite redshift at the Schwarzschild
radius, $R=2GM/c^2$, and timelike test particle geodesics can
traverse it \textit{in either direction}, so long as the initial
conditions are chosen in a manner that permits the `time'
coordinate to change in a positive sense. However, a central
singularity still exists in these coordinates and they have a
surface of infinite redshift as $r \rightarrow \infty$, at which
$ds^2 \rightarrow 0  $. This is an extreme example of a coordinate
transformation changing the size of the event horizon radius.}
Such spacetime manifolds are known
as `bundle complete' [Wheeler \& Ciuofolini 1995].

As a guiding principle, we look for solutions of the Einstein
equations
\begin{equation}
G^{\mu\nu}=(8\pi G/c^4) T^{\mu\nu}
\end{equation}
that satisfy the SPOE requirement for timelike world line
completeness. Since there is nothing in the Einstein tensor
$G^{\mu\nu}$ that enforces this condition, we must rely on
non-gravitational forces in $T^{\mu \nu}$ to dynamically enforce
it. Since the energy-momentum tensor $T^{\mu\nu}$ serves as both a
source of curvature in the Einstein equations and a generator of
the equations of motion of matter, constraints on $T^{\mu\nu}$
that enforce timelike world line completeness can also eliminate
the occurrence of event horizons. Thus the SPOE requires that the
right hand side of the GR field equation must contain
non-gravitational elements capable of stopping the collapse of
physical matter before the formation of a `trapped surface'.  This
dynamically escapes the Hawking and Penrose theorem which states
that once a trapped surface is
formed, an event horizon and curvature singularities are unavoidable.

We can show how the SPOE constrains the solutions of the Einstein
field equations. Consider a comoving interior metric given by
\begin{equation}
ds^2=A(r,t)^2c^2dt^2 - B(r,t)^2dr^2- R(r,t)^2(d\theta^2 +
sin^2\theta d\phi^2)
\end{equation}
and an exterior Vaidya metric with outgoing radiation
\begin{equation}
ds^2=(1-2GM/c^2R)c^2du^2 + 2c du dR - R^2(d\theta^2 + sin^2\theta
d\phi^2)
\end{equation}
where $R$ is the areal radius and $u=t-R/c$ is the retarded
observer time. Following Lindquist, Schwarz \& Misner [1965], we
define
\begin{equation}
\Gamma= \frac{dR}{dl}
\end{equation}
\begin{equation}
U= \frac{dR}{d\tau}
\end{equation}
\begin{equation}
M(r,t)= 4\pi \int_{0}^{r}{\rho R^2 \frac{dR}{dr} {dr}}
\end{equation}
\begin{equation}
\Gamma^2 = (\frac{dR}{dl})^2 = 1 - \frac{2GM(r,t)}{c^2R} +
\frac{U^2}{c^2}
\end{equation}
where $dl$ is a proper length element in a zero angular momentum
comoving frame, $d\tau$ an increment of proper time, $U$ is the
proper time rate of change of the radius associated with the
invariant circumference of the collapsing mass, and $M(r,t)$ is
the mass enclosed within this radius. The last two of the
relations above have been obtained from the $G_0^0$ component of
the field equation [Lindquist, Schwarz \& Misner 1965]. At the
boundary of the collapsing, radiating surface, s, we find that the
proper time will be positive definite, as required for timelike
world line completeness if
\begin{equation}
d\tau_s= \frac{du}{1+z_s} =
    du((1 - \frac{2GM(r,t)_s}{c^2R_s} + \frac{U_s^2}{c^2})^{1/2} + \frac{U_s}{c}) > 0
\end{equation}
where $z_s$ is the distantly observed redshift of the collapsing
surface. From Eq. (10) we see that in order to avoid a violation
of the requirement of timelike world line completenes for  $Us <
0$, it is necessary to dynamically enforce the `no trapped surface
condition'. \footnote {It might be argued that there might not be
a surface that physically divides matter from radiation inside a
collapsing massive continuum, however, it was first shown by Mitra
[Mitra 2000, 2002] and later corroborated by Leiter \& Robertson
[2003] that Eq.s (7 - 9) and the $G_0^0$ field equation in a zero
angular momentum comoving frame produces the `no trapped surface
condition' for any interior R(r,t). For the MECO, timelike world
line completeness is maintained by photon pressure generated by
the equipartition magnetic field everywhere in the comoving frame.
We can consider any interior location and the radiation flux there
without requiring a joined Vaidya metric. But there will
ultimately be an outer radiating boundary and the required match
to the non-singular outgoing exterior Vaidya metric guarantees
that there will be no metric singularity there.}
\begin{equation}
\frac{2GM_s}{c^2R_s} < 1
\end{equation}

\section{A Radiating, Collapsing, Magnetic Object}
The simplest form of the energy-momentum tensor that can satisfy
the SPOE requirement of timelike world line completeness, is one
that describes a collapsing, radiating plasma with an
equipartition magnetic field that emits outgoing radiation.
Between the extremes of pure magnetic energy [Thorne 1965] and
weakly magnetic, radiation dominated polytropic gases or
pressureless dust [Baumgarte \& Shapiro 2003] there are cases
where the rate of collapse can be stable. To first order, in an
Eddington limited radiation dominated context, these can be
described by the energy momentum tensor:
\begin{equation}
T_{\mu}^{\nu} = (\rho + P/c^2)u_\mu u^\nu - P \delta_\mu^\nu +
E_\mu^\nu
\end{equation}
where $E_\mu^\nu = Qk_\mu k^\nu$, $k_\mu k^\mu = 0$ describes
outgoing radiation in a geometric optics approximation, $\rho$ is
energy density of matter and $P$ the pressure.\footnote {Energy
momentum tensors corresponding to metrics describing ingoing
radiation, which are used in many black hole model calculations,
(e.g. Baumgarte \& Shapiro [2003]) cannot be used here because
they are incompatible with the $Q > 0$ boundary conditions
associated with collapsing, outwardly radiating objects.} Here $Q$
is given by
\begin{equation}
Q=\frac{-(dM/du)/4 \pi R^2}{(\Gamma_s + U_s/c)^2}
\end{equation}
At the comoving MECO surface the luminosity is
\begin{equation}
L= 4\pi R^2 Q~~ >0.
\end{equation}
and the distantly observed luminosity is
\begin{equation}
L_{\infty} = -c^2\frac{dM_s}{du} = -c^2\frac{dM_s}{d\tau(1 + z_s)}
\end{equation}
After examining the relations between surface and distantly
observed luminosities, we will use this relation to determine the
MECO lifetime.

\section{Eddington Limited MECO}
Among the various equations associated with the collapse
process there are three proper time differential equations applicable to
a compact collapsing and radiating physical surface. When evaluated on the
physical surface [Hernandez Jr.\& Misner, 1966,
Lindquist, Schwartz \& Misner 1965, Misner 1965, Lindquist,
1966] these equations are:
\begin{equation}
\frac{dU_s}{d\tau} = (\frac{\Gamma^2}{\rho+P/c^2})_s
(-\frac{\partial P}{\partial R})_s - (\frac{G(M + 4\pi R^3 (P + Q
) / c^2)}{R^2})_s
\end{equation}
\begin{equation}
\frac{dM_s}{d\tau}  =  - (4\pi R^2 P c \frac{U}{c})_s - (L (\frac{U}{c} + \Gamma))_s
\end{equation}
\begin{equation}
\frac{d\Gamma_s}{d\tau}  = \frac{G}{c^4}(\frac{L}{R})_s +
\frac{U_s}{c^2} (\frac{\Gamma^2}{\rho+ P /c^2})_s (-\frac{\partial
P}{\partial R})_s
\end{equation}
In Eddington limited steady collapse, the conditions $dU_s/d\tau
=0$ and $U_s \approx 0$ hold after some time, $\tau_{Edd}$, that
has elaspsed in reaching the Eddington limited state. Then
\begin{equation}
\frac{dU_s}{d\tau} = \frac{\Gamma_s^2}{(\rho +
P/c^2)_s}(-\frac{\partial P}{\partial R})_s
  - \frac{GM_s}{R_s^2}  = 0
\end{equation}
Where
\begin{equation}
M_s = (M + 4\pi R^3(P + Q )/c^2)_s
\end{equation}
includes the magnetic field energy in the pressure term and radiant energy in Q.

Eq. (19) when integrated over a closed surface can be solved for
the net outward flow of Eddington limited luminosity through the
surface. Taking the escape cone factor of
$27(R_g/R_s)^2/(1+z_s)^2$ into account, where $R_g=GM/c^2$, (See
Appendix A) the outflowing (but not all escaping) surface
luminosity, L, would be
\begin{equation}
L_{Edd}(outflow)_s  =\frac{4\pi G M_s c R^2(1 + z_{Edd})^3}
    {27 \kappa R_g^2}
\end{equation}
where $\kappa \approx 0.4$ cm$^2 / g$ is the plasma opacity. (For
simplicity, we have assumed here that the luminosity actually
escapes from the MECO surface rather than after conveyance through
a MECO atmosphere and photosphere. The end result is the same for
distant observers.) However the luminosity $L_s$ which appears in
Eq.s (16 - 20) is actually the net luminosity, which escapes
through the photon sphere, and is given by $L_s =
L_{Edd}(escape)_s = L_{Edd}(outflow) - L_{Edd}(fallback) =
L_{Edd,s}-L_{Edd,s}(1-27R_g^2/(R(1+z_{Edd}))^2$ Thus in Eq.s (17)
and (18), the $L_s$ appearing there is given by
\begin{equation}
L_s= L_{Edd}(escape)_s = \frac{4\pi GM(\tau)_s c
(1+z_{Edd})}{\kappa}
\end{equation}
In this context from Eq.s (9), (10), (17) and (22) we have
\begin{equation}
c^2\frac{dM_s}{d\tau} = -\frac{L_{Edd}(escape)_s}{(1+z_s)}
       =  - \frac{4\pi G M(\tau)_s c}{\kappa}
\end{equation}
which can be integrated to give
\begin{equation}
M_s(\tau) = M_s(\tau_{Edd}) \exp{((-4\pi G / \kappa c)(\tau -
\tau_{Edd}))}
\end{equation}
This yield a distantly observed MECO lifetime of $(1+z_s)\kappa
c/4\pi G \sim 5\times 10^{16}$ yr for $z_s \sim 10^8$. Finally,
equation (18) becomes
\begin{equation}
\frac{d\Gamma_s}{d\tau}
=\frac{G}{c^4}\frac{L_{Edd,s}}{R_s(\tau_{Edd})}
\end{equation}
which, in view of (13) has the solution
\begin{equation}
\Gamma_s(\tau) = \frac{1}{1 + z_s(\tau)} = (1-
\frac{2GM_s(\tau_{Edd})}{c^2R_s(\tau)_{Edd}})^{1/2}  > 0
\end{equation}
which is consistent with Eq.s (9) and (11).

If one naively attributes Eddington limit luminosity to purely
thermal processes, one quickly finds that the required MECO
surface temperatures would be so high that photon energies would
be far beyond the pair production threshhold and the compactness
would assure that photon-photon collisions would produce numerous
electron-positron pairs. Thus the MECO surface region must be
dominated by a pair plasma. Pelletier \& Marcowith [1998] have
shown that the energy density of magnetic perturbations in
equipartition pair plasmas is preferentially converted to photon
pressure, rather than causing particle acceleration. The radiative
power of an equipartition pair plasma is proportional to $B^4$,
(pair density $\propto B^2$ and synchrotron energy production
$\propto B^2$.) Lacking the equipartition pair plasma, magnetic
stress, $B^2/8\pi$, and gravitational stress, $GM\rho/R$, on mass
density $\rho$, would both increase as $R^{-4}$ during
gravitational collapse. Magnetic fields much below equipartition
levels would be incapable of stopping the collapse. However, since
photon pressure generated by the pairs at equipartition increases
more rapidly than gravitational stresses, it is possible to
stabilize the rate of collapse at an Eddington limit rate. With
this extremely efficient photon-photon pair production mechanism,
the radiation temperature and pressure is buffered near the pair
production threshold by two types of highly redshifted quantum
electrodynamic phase transitions which convert photons into pairs
on the MECO surface. The first one involves optically thick
photon-photon pair production while the second one occurs for MECO
surface magnetic fields strong enough to create bound pairs out of
the quantum electrodynamic vacuum. In the context of an Eddington
limited balance generated by the former process, the latter
process can lead to excess production of pairs, followed by excess
photon pressure and an expansion of the MECO surface. In this
manner the MECO Eddington limited collapse rate is inherently
stable (see Appendix D and E). Stability is maintained by
increased (decreased) photon pressure ($\propto B^4$) if the field
is increased (decreased) by compression or expansion. For
equipartition conditions, the field also exceeds that required to
confine the pair plasma. Since the photon luminosity is not
confined to the core it will not be trapped, as occurs with
neutrinos, however, the radiation should be thermalized as it
diffuses through an optically thick environment. \textit{To reduce
the field to the distantly observed levels implied by our analysis
of GBHC observations would require the existence of a red shift of
$z = 1.5\times 10^8(M/7M_\odot)^{1/2}$ (see Appendixes D and E)}.
The residual, distantly observable magnetic moment and extremely
faint, redshifted radiations would be the only things that would
distinguish such an object from a black hole. \footnote{An
additional point of support for very large values of redshift
concerns neutrino transport in stellar core collapse. If a
diffusion limited neutrino luminosity of $\sim 10^{52}$ erg/s
[Shapiro \& Teukolsky 1983] were capable of very briefly
sustaining a neutrino Eddington limit rate of collapse, then the
subsequent reduction of neutrino luminosity as neutrino emissions
are depleted and trapped in the core would lead to an adiabatic
collapse, magnetic flux compression, and photon emissions reaching
an Eddington limit. At this point the photon luminosity would need
to support a smaller diameter and more tightly gravitationally
bound mass. A new photon Eddington balance would thus require an
escaping luminosity reduced by at least the $\sim 10^{20}$ opacity
ratio $(\sigma_T/\sigma_\nu)$, where $\sigma_T = 6.6\times
10^{-25}$ cm$^2$ is the Thompson cross section and $\sigma_\nu =
4.4\times 10^{-45}$ cm$^2$ is the neutrino scattering
cross-section. Thus $L_\infty <10^{31-32}$ erg/s would be
required. For this to correspond to an Eddington limit luminosity
as distantly observed would require $1+z \sim 10^8$. The adiabatic
relaxation of neutrino support and formation of a pair plasma is
an important step in gravitational collapse that is not
encompassed by polytropic equation of state models of collapse. It
is of some interest that if neutrinos have non-zero rest mass they
might be trapped inside the photon sphere anyway.}

An electron-positron pair atmosphere of a MECO is an extremely
significant structure that conveys radiation from the MECO surface
to a zone with a much lower red shift and larger escape cone from
which it escapes.  In order to describe this process
computationally within a numerical grid, a radial grid interval no
larger than $\sim 10^{-8}R_g$ would be needed, where $R_g =
GM/c^2$ is the gravitational radius. Although there have been many
numerical studies of the behavior of collapsing compact objects in
GR, to our knowledge none have sufficient numerical resolution to
examine the extreme red shift regime associated with MECO nor have
they considered the emergent properties of equipartition magnetic
fields and pair plasmas at high red shift. Until computer models
of gravitational collapse encompass these crucial physical and
computational elements, simulations that apparently produce black
hole states must be regarded as mere speculations.

\section{The Quiescent MECO}
The quiescent luminosity of a MECO originates deep within its photon sphere.
When distantly observed it is diminished by both gravitational red shift
and a narrow exit cone. The gravitational red shift
reduces the surface luminosity by $1/(1+z)^2$ while the exit cone further
reduces the luminosity by the factor
$27 R_g^2/(R(1+z))^2 \sim 27/(4(1+z)^2)$ for large z. (See Appendix A).
Here we have used
\begin{equation}
\frac{R_g}{R} = \frac{1}{2}(1 - \frac{1}{(1+z)^2})~< \frac{1}{2}
\end{equation}
where R and z refer to the location from which photons escape.
The net outflow fraction of the luminosity  provides the support for
the collapsing matter, thereby dynamically maintaining the SPOE requirement
of timelike world line completeness. The photons which finally escape do so
from the photosphere of the pair atmosphere. The fraction of
luminosity from the MECO surface
that escapes to infinity in Eddington balance is
\begin{equation}
(L_{Edd})_s = \frac {4\pi G M_s c(1+z)}{\kappa} = 1.27 \times
10^{38}m(1+z_s)~~~~ erg/s
\end{equation}
where $m=M/M_\odot$. The distantly observed luminosity is:
\begin{equation}
L_\infty = \frac{(L_{Edd})_s}{(1+z_s)^2} = \frac {4\pi G M_s
c}{\kappa(1+z_s)}
\end{equation}
When radiation reaches the photosphere, where the temperature is $T_p$,
the fraction that escapes to be distantly observed is:
\begin{equation}
L_\infty = \frac{4 \pi R_g^2 \sigma T_p^4 27}{(1+z_p)^4}
    = 1.56\times 10^7 m^2 T_p^4 \frac{27}{(1+z_p)^4}~~~~erg/s
\end{equation}
where $\sigma = 5.67\times 10^{-5}$ erg/s/cm$^2$ and subscript p
refers to conditions at the photosphere. Eq.s (29) and (30) yield:
\begin{equation}
T_\infty = T_p/(1+z_p) = \frac{2.3\times 10^7}{(m(1+z_s))^{1/4}}~~~~ K.
\end{equation}
To examine typical cases, a $10 M_\odot$, $m = 10$ GBHC modeled in
terms of a MECO with $z = 1.5\times 10^8(m/7)^{1/2}$ would have
$T_\infty = 1.1\times 10^5 K = 0.01$ keV, a bolometric luminosity,
excluding spin-down contributions, of $L_\infty =7.3\times 10^{30}
erg/s$, and a spectral peak at 220 A$^0$, in the photoelectrically
absorbed deep UV. For an m=$10^7$ AGN, $T_\infty = 630 K$,
$L_\infty = 7.2\times 10^{33} erg/s$ and a spectral peak in the
infrared at 4$\mu$. (Sgr A$^*$ at $m \approx 3\times 10^6$, would
have $T_\infty =1100$ K, and a 2.2 micron brightness below $0.6$
mJy; more than an order of magnitude below the observational upper
limit of 9 mJy [Reid et al. 2003].) Hence passive MECO without
active accretion disks, although not black holes, have lifetimes
much greater than a Hubble time and emit highly red shifted
quiescent thermal spectra that may be quite difficult to observe.
There are additional power law components of similar magnitude
that originate as magnetic dipole spin-down radiation
(Table 1 and see below).

Escaping radiation passes through a pair plasma atmosphere that
can be shown, \textit{ex post facto} (See Appendix B), to be
radiation dominated throughout. In fact, at the extreme redshifts
contemplated here, Mitra (2006) has shown that the interior of the
collapsed object must be radiation dominated. Under these
circumstances, the radiation pressure within the equilibrium
atmosphere obeys $P_{rad}/(1+z) = constant$\footnote{Due to its
negligible mass, we consider the pair atmosphere to exist external
to the Meco. Due to the slow collapse, the exterior Vaidya metric
can be approximated by exterior, outgoing Finkelstein coordinates.
In this case, the hydrostatic balance equation within the MECO
atmosphere is $\frac{\partial p}{\partial r} = -\frac{\partial
\ln{(g_{00})}}{2
\partial r}(p + \rho c^2)$, where $g_{00} = (1-2R_g/r)$ and $\rho
c^2 << p$. This integrates to $p/(1+z) = constant$.}. Thus the
relation between surface and photosphere temperatures is
$T_s^4/(1+z_s) = T_p^4/(1+z_p)$. At the MECO surface, we expect a
pair plasma temperature of $T_s \approx m_ec^2/k \sim 6\times 10^9
K$ because an equipartition magnetic field effectively acts as a
thermostat which buffers the temperature of the optically thick
synchrotron radiation escaping from the MECO surface [Pelletier \&
Marcowith 1998]. But since $T_\infty = T_p/(1+z_p)$, and using
$T_s=6\times 10^9$ K, we have that
\begin{equation}
T_p = T_s(\frac{T_s}{T_\infty (1+z_s)})^{1/3} = 3.8\times
10^{10}\frac{m^{1/12}}{(1+z_s)^{1/4}}=4.5\times 10^8
(m/7)^{-1/24}~K
\end{equation}
In the last expression, we have used $1+z_s=1.5\times
10^8(m/7)^{1/2}$. Using Eq. (31), this leads immediately to
$(1+z_p)=3500\times (m/7)^{1/3}$, independent of the surface
redshift, thus confirming that for MECO with pair atmospheres to
exist, they must be inherently highly redshifted. Due to the very
weak dependence of $T_p$ on $m$, the photosphere temperatures of
MECO are all very nearly $4.5\times 10^8$ K.

\section{An Actively Accreting MECO}
From the viewpoint of a distant observer, accretion would
deliver mass-energy to the\linebreak MECO, which would then radiate most of it
away. The contribution from the central MECO alone would be
\begin{equation}
L_\infty = \frac {4\pi G M_s c}{\kappa(1+z_s)}+ \frac{\dot{m}_\infty c^2}{1+z_s}(e(1+z_s)-1)
    = 4 \pi R_g^2 \sigma T_p^4 \frac {27} {(1+z_p)^4}
\end{equation}
where $e = E/mc^2 = 0.943$ is the specific energy per particle
available after accretion disk flow to the marginally stable orbit radius,
$r_{ms}$. Assuming that $\dot{m}_\infty$ is some fraction, f, of the
Newtonian Eddington limit mass accretion rate, $4\pi G M c/\kappa$, then
\begin{equation}
1.27\times 10^{38}\frac{m\eta}{1+z_s} =
(27)(1.56\times 10^7)m^2(\frac{T_p}{1+z_p})^4
\end{equation}
where $\eta=1+f((1+z_s)e -1)$ includes both quiescent and
accretion contributions to the luminosity. Due to the extremely
strong dependence on temperature of the density of pairs, (see
Appendix B) it is unlikely that the temperature of the photosphere
will be greatly different from the average of $4.6\times 10^8 K$
found previously for a typical GBHC. Assuming this to be the case,
along with $z=10^8$, $m=10$, and $f=1$, we find $T_\infty =
T_p/(1+z_p) = 1.3\times 10^7 K$ and $(1+z_p) = 35$, which
indicates considerable photospheric expansion. The MECO luminosity
would be approximately Newtonian Eddington limit at $L_\infty =
1.2\times 10^{39}$ erg/s. For comparison, the accretion disk
outside the marginally stable orbit at $r_{ms}$ (efficiency =
0.057) would produce only $6.8\times 10^{37}$ erg/s. Thus the high
accretion state luminosity of a GBHC would originate primarily
from the central MECO. The thermal component would be `ultrasoft'
with a temperature of only $1.3\times 10^7 K$ (1.1 keV). A
substantial fraction of the softer thermal luminosity would be
Compton scattered to higher energy in the plunging flow inside
$r_{ms}$. Even if a disk flow could be maintained all the way to
the MECO surface, where a hot equatorial band might result, the
escaping radiation would be spread over the larger area of the
photosphere due
to photons origins deep inside the photon orbit.

For radiation passing through the photosphere most photons would
depart with some azimuthal momentum on spiral trajectories that
would eventually take them across and through the accretion disk.
Thus a very large fraction of the soft photons would be subject to
bulk comptonization in the plunging region inside $r_{ms}$. This
contrasts sharply with the situation for neutron stars where there
probably is no comparable plunging region and few photons from the
surface cross the disk. This could account for the fact that hard
x-ray spectral tails are comparatively much stronger for high
state GBHC. Our preliminary calculations for photon trajectories
randomly directed upon leaving the photon sphere indicate that
this process would produce a power law component with photon index
greater than 2. These are difficult, but important calculations
for which the effects of multiple scattering must be considered.
But they are beyond the scope of this work, which is intended as a
first description of the
general MECO model.

\section{Magnetosphere - Disk Interaction}
In LMXB, when the inner disk engages the magnetosphere, the inner
disk temperature is generally high enough to produce a very
diamagnetic plasma. This may not be the case for AGN. Surface
currents on the inner disk distort the magnetopause and they also
substantially shield the outer disk such that the region of strong
disk-magnetosphere interaction is mostly confined to a ring or
torus, of width $\delta r$ and half height $H$. This shielding
leaves most of the disk under the influence of its own internal
shear dynamo fields, [e.g. Balbus \& Hawley 1998, Balbus 2003]. At
the inner disk radius the magnetic field of the central MECO is
much stronger than the shear dynamo field generated within the
inner accretion disk. In MHD approximation, the force density on
the inner ring is $F_v = (\nabla \times B) \times B / 4\pi$. For
simplicity, we assume coincident magnetic and spin axes of the
central object and take this axis as the $z$ axis of cylindrical
coordinates $(r,\phi,z)$.

The magnetic torque per unit volume of plasma in the inner ring of
the disk that is threaded by the intrinsic magnetic field of the
central object, can be approximated by $\tau_v=rF_{v\phi} = r
\frac{B_z}{4\pi} \frac{\partial B_{\phi}}{\partial z} \sim r
\frac{B_zB_{\phi}}{4\pi H}$, where $B_{\phi}$ is the average
azimuthal magnetic field component. We stress that $B_{\phi}$, as
used here, is an average toroidal magnetic field component. The
toroidal component likely varies episodically between reconnection
events [Goodson \& Winglee 1999, Matt et al. 2002, Kato, Hayashi
\& Matsumato 2004, Uzdensky 2002].

The average flow of disk angular momentum entering the inner ring
is $\dot{M}r v_k$, where $\dot{M}$ is mass accretion rate and
$v_k$ is the Keplerian speed in the disk. This angular momentum
must be extracted by the magnetic torque, $\tau$, hence:
\begin{equation}
\tau = \dot{M}r v_k = r\frac{B_zB_{\phi}}{4\pi H}(4\pi r H \delta
r).
\end{equation}
In order to proceed further, we assume that $B_{\phi} = \lambda
B_z$, $B_z=\mu/r^3$, and use $v_k=\sqrt{GM/r}$, where $\lambda$ is
a constant, presumed to be of order unity, $\mu$ is the magnetic
dipole moment of the central object $M$, its mass, and $G$, the
Newtonian gravitational force constant. With these assumptions we
obtain
\begin{equation}
\dot{M} = (\frac{\lambda \delta r}{r}) \frac {\mu^2}{r^5 \omega_k}
\end{equation}
where $\omega_k = v_k/r$ and the magnetopause radius, $r_m$ is
given by
\begin{equation}
r_m = (\frac{\lambda \delta
r}{r})^{2/7}(\frac{\mu^4}{GM\dot{M}^2})^{1/7}
\end{equation}

In order to estimate the size of the boundary region, $(\delta r
/r)$, we normalized this disk-magnetosphere model for agreement
with  radii calculated for an elaborate model of a gas pressure
dominated disk [Ghosh \& Lamb 1992]. Although we find the portion
of the inner disk threaded by magnetic fields to be smaller than
the Ghosh \& Lamb model, this size for the inner radius yields
very accurate results for accreting millisecond pulsars, which
have known magnetic moments. We find $(\frac{\lambda \delta r}{r})
= 0.015$, which indicates a gratifyingly small strong interaction
zone for disk and magnetosphere. Using units of $10^{27}$ gauss
cm$^3$ for magnetic moments, $100$ Hz for spin, $10^6$ cm for
radii, $10^{15}$ g/s for accretion rates, solar mass units,
$\lambda \delta r/r = 0.015$ and otherwise obvious notation, we
find the magnetosphere radius to be:
\begin{equation}
r_m~=~8\times 10^6 {(\frac{\mu_{27}^4}{m
\dot{m}_{15}^2})}^{1/7}~~~~ cm
\end{equation}
where $m=M/M_\odot$ and the disk luminosity is
\begin{equation}
L = \frac{GM\dot{m}}{2r_m}
\end{equation}
The co-rotation radius, at which disk Keplerian and magnetosphere
spins match is:
\begin{equation}
r_c = 7\times 10^6{(\frac{m}{\nu_2^2})^{1/3}} ~~~cm
\end{equation}
The low state disk luminosity at the co-rotation radius is the
maximum luminosity of the true low state and is given by:
\begin{equation}
L_c = \frac{GM\dot{m}}{2r_c}=1.5 \times 10^{34} \mu_{27}^2
{\nu_2}^3 m^{-1}~~~~erg/s
\end{equation}
The minimum high state luminosity for all accreting matter being
able to reach the central object occurs at approximately the same
accretion rate as for $L_c$ and is given by:
\begin{equation}
L_{min} = \xi \dot{m}c^2 = 1.4\times 10^{36} \xi \mu_{27}^2
\nu_2^{7/3} m^{-5/3}~~~erg/s
\end{equation}
Where $\xi \sim 0.42$ for MECO for the photon sphere\footnote{The
time for a luminosity variation to be observed is very long for
energy released by processes inside the photon sphere.} and $\xi =
0.14$ for NS is the efficiency of accretion to the central
surface.

In true quiescence, the inner disk radius is larger than the light
cylinder radius. In NS and GBHC, the inner disk may be ablated due
to radiation from the central object. The inner disk radius can be
ablated to distances larger than $5\times 10^4 km$ because
optically thick material can be heated to $\sim 5000 K$ and
ionized by the radiation. The maximum disk luminosity of the true
quiescent state occurs with the inner disk radius at the light
cylinder, $r_{lc}=c/\omega_s= r_m$. The maximum luminosity of the
quiescent state is typically a factor of a few larger than the
average observed quiescent luminosity.
\begin{equation}
L_{q,max} = (2.7\times
10^{30} erg/s) \mu_{27}^2 \nu_2^{9/2} m^{1/2}
\end{equation}

We calculate the average quiescent luminosities in the soft x-ray
band from $\sim 0.5 - 10$ keV using the correlations of Possenti
et al. [2002] with spin-down energy loss rate as:
\begin{equation}
L_q = \beta \dot{E} = \beta 4 \pi^2 I \nu \dot{\nu}
\end{equation}
where $I$ is the moment of inertia of the star, $\nu$ its rate of
spin and $\beta$ a multiplier that can be determined from this new
$\dot{E} - L_q$ correlation for given $\dot{E}$; i.e., known spin
and magnetic moment. In previous work we had used $\beta =
10^{-3}$ for all objects, but $\beta \sim 3\times 10^{-4}$ would
be the average value for GBHC-MECO consistent with the Possenti
correlation. We assume that the luminosity is that of a spinning
magnetic dipole for which $\dot{E} = 32\pi^4 \mu^2 \nu^4/3c^3$,
(Bhattacharya \& Srinivasan 1995] where $\mu$ is the magnetic
moment. Thus the quiescent x-ray luminosity would then be given
by:
\begin{equation}
L_q = \beta \times \frac{32 \pi^4 \mu^2 \nu^4}{3c^3} = 3.8 \times
10^{33} \beta \mu_{27}^2 \nu_2^4 ~~~~~erg/s
\end{equation}

According to the Possenti correlation, $\beta = L_q/\dot{E}~~
\propto~ \dot{E}^{0.31}$. $\beta$ should be a dimensionless,
ratio, and independent of mass. But since $\dot{E}$ is
proportional to mass, we extend the Possenti relation, without
loss of generality, to provide a mass scale invariant quantity. We
therefore take $\beta \propto (\dot{E}/m)^{0.31}$. From the
Possenti correlation, assuming all the objects in their study have
the canonical $m=1.4$, we then find that
\begin{equation}
\beta=7\times 10^{-4}(\dot{E}/m)^{0.31}=4.6\times
10^{-4}(10^{-36}L_c\nu_2)^{0.31}
\end{equation}

Since the magnetic moment, $\mu_{27}$, enters each of the above
luminosity equations it can be eliminated from ratios of these
luminosities, leaving relations involving only masses and spins.
For known masses, the ratios then yield the spins. Alternatively,
if the spin is known from burst oscillations, pulses or spectral
fit determinations of $r_c$, one only needs one measured
luminosity, $L_c$ or $L_{min}$ at the end of the transition into
the soft state, to enable calculation of the remaining $\mu_{27}$
and $L_q$. For some GBHC, we found it to be necessary to estimate
the co-rotation radius from multicolor disk fits to the thermal
component of low state spectra. The reason for this is that the
luminosities are sometimes unavailable across the whole spectral
hardening transition from $L_c$ to $L_{min}$ for GBHC. 

For GBHC, it is a common finding that the low state inner disk
radius is much larger than that of the marginally stable orbit;
e.g. [Markoff, Falcke \& Fender 2001, $\dot{Z}$ycki, Done \& Smith
1997a,b 1998, Done \& $\dot{Z}$ycki 1999, Wilson \& Done 2001].
The presence of a magnetosphere is an obvious explanation. Given
an inner disk radius at the spectral state transition, the GBHC
spin frequency follows from the Kepler
relation $2 \pi \nu_s = \sqrt{GM/r^3}$.

Although we have taken our model and used it to predict the spin
rates and accurate quiescent luminosities for NS  and GBHC that
are shown in Table 1, it now appears that we could use the fact
that the model fits well to calculate more accurate parameters. By
placing the last mass scale invariant Possenti relation for
$\beta$ into the relation for quiescent luminosity and using it
with the expression for $L_c$, we can determine spin rates to be
given by
\begin{equation}
\nu = 89(L_{q,32}/(mL_{c,36}^{1.31}))^{1/1.31}~~~Hz
\end{equation}
where $L_{q,32} = 10^{-32}L_q$ and $L_{c,36} = 10^{-36}L_c$. Using
this relation and the equation for $L_c$, the average spin rate
for the GBHC of Table 1 is reduced to 10 Hz and the average GBHC
magnetic moment is found to be 2200 gauss-cm$^3$. These results
essentially preserve the quiescent luminosities and actually
should be more reliably determined because luminosity measurements
are more reliable than inner disk radii determined from spectral
fitting.\footnote{MECO Magnetic moments must scale as $m^{5/2}$.
Then for consistent change of results reported in Table 1, this
requires $L_{c,36}/(m^4 \nu^3) =1$.}

\section{ Low State Mass Ejection and Radio Emission}
The radio flux, $F_{\nu}$, of jet sources has a power law
dependence on frequency of the form
\begin{equation}
F_{\nu}~ \propto~ \nu^{-\alpha}
\end{equation}
It is believed to originate in jet outflows and has been shown to
be correlated with the low state x-ray luminosity [Merloni, Heinz
\& DiMatteo 2003], with $F_\nu \sim L_x^{0.7}$.  The radio
luminosity of a jet is a function of the rate at which the
magnetosphere can do work on the inner ring of the disk. This
depends on the relative speed between the magnetosphere and the
inner disk; i.e., $\dot{E} =\tau (\omega_s - \omega_k)$, or
\begin{equation}
\dot{E} = (\frac{\lambda \delta r}{r})\frac{\mu^2 \omega_s (1 -
\frac{\omega_k}{\omega_s})}{r^3}
    ~\propto~ \mu^2 M^{-3}\dot{m}_{Edd}^{6/7}\omega_s(1 - \frac{\omega_k}{\omega_s})
\end{equation}
Here $\dot{m}_{Edd}$ is the mass accretion rate divided by the
rate that
would produce luminosity at the Eddington limit for mass $M$.

Disk mass, spiraling in quasi-Keplerian orbits from negligible
speed at radial infinity must regain at least as much energy as
was radiated away in order to escape. For this to be provided by
the magnetosphere requires $\dot{E} \geq GM\dot{M}/2r$, from which
$\omega_k \leq 2\omega_s/3$. Thus the magnetosphere alone is
incapable of completely ejecting all of the accreting matter once
the inner disk reaches this limit and the radio luminosity will be
commensurately reduced and ultimately cut off at maximum x-ray
luminosity for the low state and $\omega_k=\omega_s$. Typical data
for GX339-4 [Gallo, Fender \& Pooley 2003] are shown in Figure 1.
For very rapid inner disk transit through the co-rotation radius,
fast relative motion between inner disk and magnetosphere can heat
the inner disk plasma and strong bursts of radiation pressure from
the central object may help to drive large outflows while an
extended jet structure is still largely intact. This process has
been calculated \footnote{though for inner disk radii inside the
marginally stable orbit [Chou \& Tajima 1999]}using pressures and
poloidal magnetic fields of unspecified origins. A MECO is
obviously capable of supplying both the field and a radiation
pressure. The hysteresis of the low/high and high/low state
transitions may be associated with the need for the inner disk to
be completely beyond the corotation radius before a jet can be
regenerated
after it has subsided.
%
%
%
\begin{figure}
\epsfig{figure=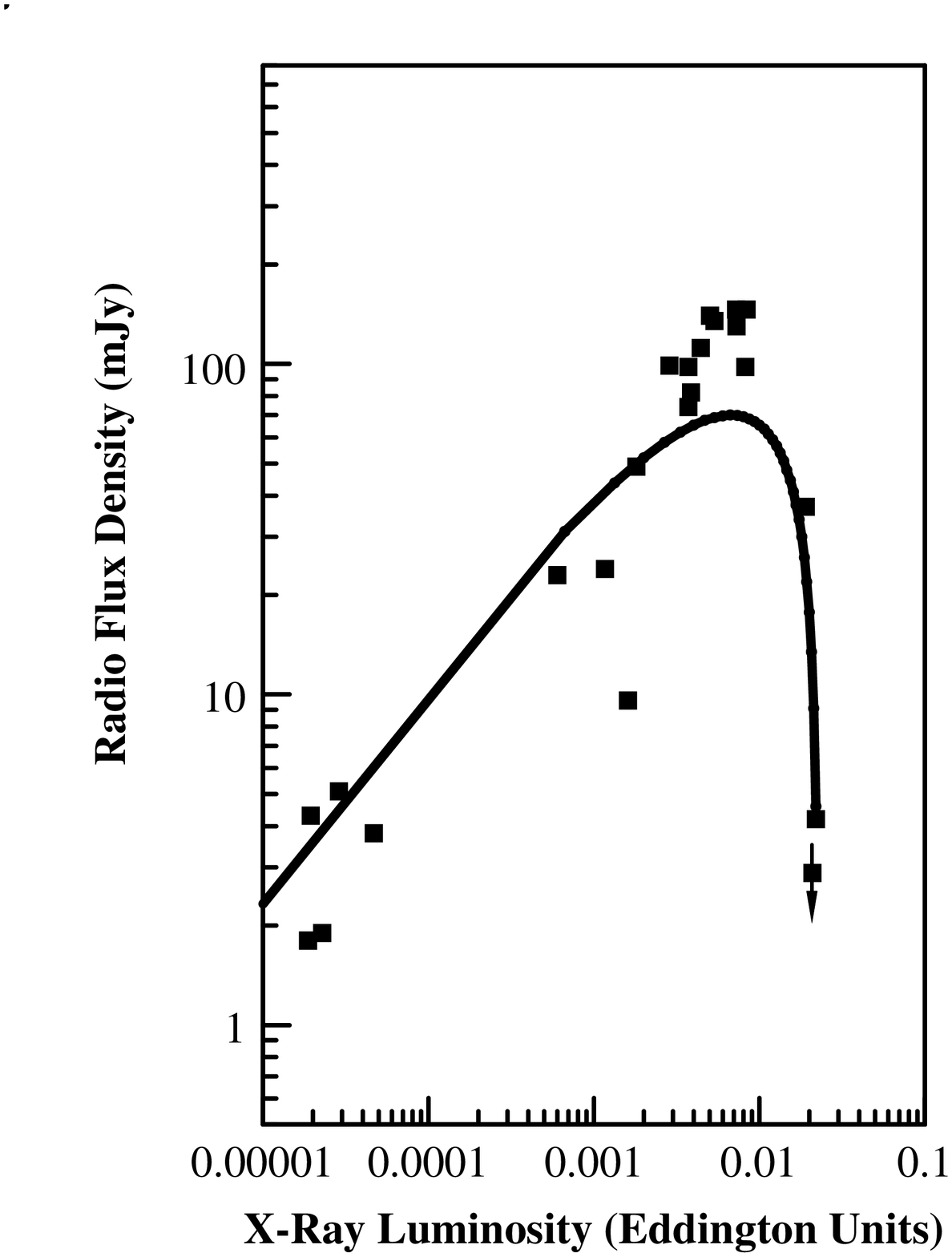,angle=0,width=12cm} \caption
{Radio/X-ray correlation for GX 339-4. Data from [Gallo, Fender \&
Pooley 2003]. The line is from Eq. (51) with $L_c$ from Table 1,
and $C(M,\beta,\omega_s)$ crudely estimated. The line illustrates
the predicted radio cutoff of the MECO model.}
\end{figure}%

Since $\dot{E} \propto r^{-3}$ and $L_d \propto r^{-9/2}$, it is
apparent that we should expect radio luminosity, $L_R \propto
L_d^{2/3}$. In particular we find
\begin{equation}
L_R = C(M,\beta, \omega_s)2L_c^{1/3}
L_d^{2/3}(1-\omega_k/\omega_s)
\end{equation}
where $\beta =\mu/M^3$ and $C(M,\beta, \omega_s)$ is a constant,
dependent on the radio bandpass. It has been analyzed and
evaluated [Robertson \& Leiter 2004]. The cutoff at
$\omega_k=\omega_s$ is shown by the line on Figure 1. The cutoff
typically occurs with x-ray luminosity of $\sim 0.01 - 0.02$ times
Eddington luminosity. If we let $x=L_d/L_c$, then for $x < 1$,
corresponding to the low state, Eq. (50) takes the form:
\begin{equation}
L_R =C(M,\beta, \omega_s) 2L_c(x^{2/3} - x)
\end{equation}
The function has a maximum value of $0.3C(M,\beta, \omega_s)
L_c$ at $x=0.3$.

Strictly speaking, $L_d$, in Eq. (50) should be the bolometric
luminosity of the disk, however, the x-ray luminosity over a large
energy band is a very substantial fraction of the disk luminosity.
To compare with the correlation exponent of 2/3 obtained here,
recent studies, including noisy data for both GBHC and AGN have
yielded $0.71 \pm 0.01$ [Gallo, Fender \& Pooley 2003], 0.72
[Markoff et al. 2003, Falcke, K\"{o}rding \& Markoff 2003], $0.60
\pm 0.11$ [Merloni, Heinz \& Di Matteo 2003] and $0.64 \pm 0.09$
[Maccarone, Gallo \& Fender 2003]. For $\alpha$ in the range (0 to
-0.5), $\beta \propto M^{-1/2}$, $\omega_s \propto M^{-1}$ and
$L_c \propto M$, the MECO model yields $C(M) \propto
M^{(9-4\alpha)/12}$ and (neglecting the cutoff region)
\begin{equation}
log L_R = (2/3)log L_x + (0.75 - 0.92) log M + const.
\end{equation} which is a better fit to the ``fundamental plane" of
Merloni, Heinz \& Di Matteo [2003] than any of the ADAF,
disk/corona or disk/jet models they considered (see their
Figure 5 for a $\chi^2$ density plot). This last relation correctly
describes the correlation for both GBHC and AGN.

\section{Spectral States}
The progression of configurations of accretion disk, magnetic
field and boundary layer is shown for GBHC in Figure 2. The
caption summarizes the spectral features expected in four regimes:
\begin{figure}
\epsfig{figure=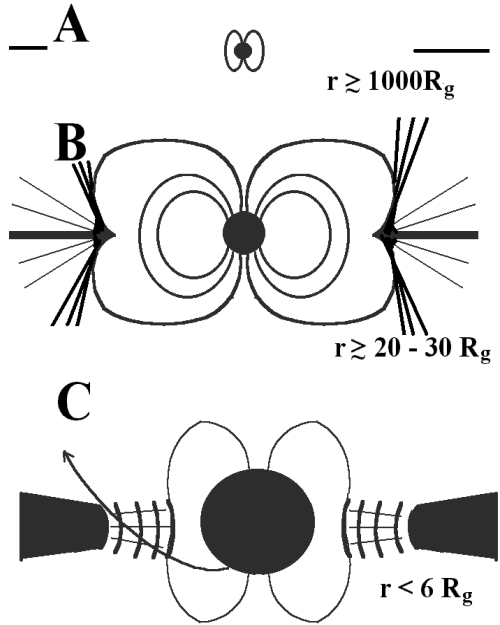,angle=0,width=8cm} \caption{MECO
Spectral States: {\bf A quiescent:} Inner disk ablated, low
accretion rate to inner ablation radius $\sim 10^9 - 10^{10} cm$
generates optical emissions. Magnetic spin-down drives hard
power-law x-ray spectrum. For NS, surface x-ray emissions may be
visible. {\bf B. Low state:} Thin, gas pressure dominated inner
disk has large magnetically dominated viscosity. The inner disk
radius lies between the light cylinder and co-rotation radii. Disk
winds and jets are driven by the magnetic propeller. A hard
spectrum is produced as most soft x-ray photons from the disk are
Comptonized by either outflow or corona. Outflows of electrons on
open magnetic field lines, possibly in jets, produce synchrotron
radiation. Most of the outer disk is shielded from the magnetic
field of the central object as surface currents in the inner disk
change the topology of the magnetopause. {\bf C: Intermediate and
High state:} Once the inner disk is inside the co-rotation radius,
the outflow and synchrotron emissions subside, but a steep power
law spectrum is produced until the jet structure dies and an
optically thick disk builds in to the marginally stable orbit.
Relaxation oscillations may occur if radiation from the central
object momentarily drives the inner disk back outside the
co-rotation radius. A boundary layer of material beginning to
co-rotate with the magnetosphere may push the magnetopause to the
star surface for NS or inside $r_{ms}$ for MECO, where a
supersonic flow plunges inward until radiation pressure stabilizes
the magnetopause or plasma interchange instabilities break up the
flow. The MECO photosphere radiates a bright `ultrasoft' thermal
component. Bulk comptonization of many photons on spiral
trajectories crossing the plunging zone inside $r_{ms}$ produces a
hard x-ray spectral tail. Declining phase hysteresis occurs since
the soft luminosity can persist until the jet structure is
rebuilt.}
\end{figure}%
{\bf Quiescence}\\
In true quiescence, the inner disk radius is outside the light
cylinder. In fact, it is usually far beyond the light cylinder, as
the inner disk is ablated by $\sim 10^{30-33}$ erg/s radiation
from the central object. This luminosity is sufficient to raise
the temperature of the optically thick inner disk above the $\sim
5000$ K instability temperature for hydrogen out to a distance of
$r \sim 10^{10}$ cm. Therefore we expect the quiescent inner disk
to be essentially empty. The rate of mass flow from ablation at
the inner disk radius would only need to be $\sim 10^{13}$ g/s to
produce the quiescent optical emission observed for GBHC and NS.
The ablated material could escape if it reached the magnetic
propeller region, which is confined to the light cylinder at a
much smaller radius, $r_{lc}$, than that of the inner disk. The
ejection of the ablated material probably also produces the
quiescent state power law x-ray spectrum. It likely would also
produce weak radio emissions, but in this case the exponent of the
radio/x-ray luminosity correlation would change from $2/3$ to $1$
as both would originate as optically thin synchrotron emissions.

The empty inner disk makes the MECO model compatible with the disk
instability model of x-ray nova outbursts, which begin as
`outside-in' events in which substantial outer mass reservoirs
have been observed to fill an accretion disk on the viscous
timescale of a very subsonic radial flow [Orosz et al. 1997]. From
true quiescence to the light cylinder, the x-ray luminosity
changes by a factor of only a few. The accretion rate at
$L_{q,max}$ is only
$\dot{m}=1.7\times 10^{12}\mu_{27}\nu_2^{7/2}m^{-1/2}~\sim 2\times 10^{14}~g/s$.

Quiescent luminosities that are generally 10 - 100 X lower for
GBHC than for NS have been claimed as evidence for the existence
of event horizons. [Narayan et al. 1997, Garcia et al. 2001]. In
the MECO model, the quiescent luminosity is driven by the magnetic
dipole radiation from the spinning central magnetic moment. The
lower quiescent luminosities of the GBHC are explained by their
lower spin rates and (perhaps unobservably) low
rates of quiescent emission from the central MECO.

{\bf Low/Hard State}\\
In the low state, the inner disk radius is inside the light
cylinder, with hot, diamagnetic plasma reshaping the magnetopause
topology [Arons et al. 1984]. This magnetic propeller regime
(Ilarianov \& Sunyaev 1975, Stella, White \& Rosner 1986, Cui
1997, Zhang, Yu \& Zhang 1997, Campana et al. 1998] exists until
the inner disk pushes inside the co-rotation radius, $r_c$. From
$r_{lc}$ to $r_c$, the x-ray luminosity may increase by a factor
of $\sim 10^3 - 10^6$. Inside $r_c$, large fractions of the
accreting plasma can continue on to the central object and produce
a spectral state switch to softer emissions. We have shown
[Section 8 and Robertson \& Leiter 2002] that magnetic moments and
spin rates can be determined from luminosities at the end points
of the transition from low/hard to high/soft spectral states. The
magnetic moments and spins were used to calculate the $\sim
10^{3-6}$ times fainter quiescent luminosities expected from
spin-down. The results are recapitulated and extended in Table 1
and Appendix C. During waning phases of nova outbursts, $L_c \sim
0.02L_{Edd}$ can be identified as the maximum disk luminosity
upon entering the low state.

Until the inner disk reaches $r_c$, accreting plasma is ejected.
It may depart in a jet, or as an outflow back over the disk as
plasma is accelerated on outwardly curved or open magnetic field
lines. Radio images of both flows have been seen [Paragi et al.
2002]. Equatorial outflows could contribute to the low state hard
spectrum by bulk Comptonization of soft photons in the outflow,
however, we think that the hard spectrum originates primarily in
patchy coronal flares [Merloni \& Fabian 2002] on a conventional
geometrically thin, optically thick disk.\footnote{See the disk
characteristics section below. Optically thick material is needed
to produce pressures capable of countering magnetic pressure on
the inner disk.} Both outflow comptonization and coronal flares
are compatible with partial covering models for dipping sources,
in which the hard spectral region seems to be extended [Church
2001, Church \& Balucinska-Church 2001]. Alternatively, a compact
jet [Corbel \& Fender 2002] might be a major contributor to the
hard spectrum, but if so, the x-ray luminosity must fortuitously
match the power that would be dissipated in a conventional thin
disk. Finally, we note that the power law emissions of the
low/hard state are usually cut off below $\sim 100$ keV,
consistent with a coronal temperature of $20 - 50$ keV. Bulk
comptonization would be expected to produce higher energies.

{\bf Intermediate (Steep Power-Law) State}\\
Intermediate states typically occur with luminosities in the range
$(0.01 - 0.3)L_{Edd}$ when some, but not all, of the accreting
matter can make its way to the central object. These states are
usually observed with rising luminosity and often do not appear
during declining phases. They are characterized by increasing
luminosity, an increasing power law index and the presence of a
weak, soft thermal contribution to the x-ray spectrum. The soft
emissions originate, at least in part from the central object
surface. The steep power law, extending well beyond $100$ keV, is
produced via bulk comptonization as soft photons scatter from the
(initially) optically thin material entering the magnetosphere.
Incomplete spectral state switches terminating well below the
Eddington limit, such as those exhibited by Cygnus X-1 may occur.
With the inner disk radius large ($>~20 R_g$) near co-rotation,
there is a very large difference in the efficiency of energy
release at the central object vs. the disk. Thus changes of
luminosity and an apparent, but incomplete, spectral state switch
can occur for very small change of accretion rate. Relaxation
oscillations between hard and soft states, driven by intermittent
radiation from the central object, can occur if the accretion rate
is not steady. Large periodic jet ejections may be associated with
this state, for which significant toroidal winding of the poloidal
magnetic field lines and radiation pressure may contribute to the
ejection. The intermediate state terminates in the high/soft state
with the disk becoming optically thick all the way into the
marginally stable orbit or a NS surface. This occurs at $\sim (0.2
- 0.4)L_{Edd}$. The outflow/jet of the low state is a substantial
flow structure that is systematically disrupted as flow in the
inner disk increases in the intermediate state.

{\bf High/Soft (Thermally Dominant) State}\\
With the disk inner radius inside $r_c$, the propeller regime ends
and matter of sufficient pressure can make its way inward. With
inner disk inside $r_c$, the outflow and/or jets subside, the
system becomes radio quiet, and a soft thermal excess from the
central object appears, [e.g., see Fig. 3.3 of Tanaka \& Lewin, p.
140], which may be even be described as `ultrasoft' [White \&
Marshall 1984]; particularly when the central object cools as the
luminosity finally begins to decline\footnote{Due to the high
redshift, the MECO luminosity decay can be very slow.}. We have
shown that GBHC-MECO would produce a dominant `ultrasoft'
component at $\sim 1.1$ keV in the high state. They would also
continue to produce a steep power-law hard tail as soft photons
leaving the MECO well inside the photon orbit take trajectories
that take them across the plunging region inside the marginally
stable orbit. Since there is no comparably rich source of photons
on disk crossing trajectories for NS and a much smaller, if any,
plunging region, there is no comparable hard spectral component
produced by bulk comptonization in their high states.

In producing the the high/soft (thermally dominant) state the jet
structure of the\linebreak low/hard state is destroyed and an optically
thick disk progressively pushes inside $r_c$, shredding the
magnetic field lines as it encroaches. The flow into the
magnetosphere is initially optically thin, but eventually gives
way to an optically thick disk that reaches the marginally stable
orbit (or star surface in NS systems). The steep power law
spectrum becomes increasingly dominant and then its luminosity may
decline (while photon energies increase) as the optically thick
part of the disk encroaches. After the soft state has peaked and
begun to decline, the flow remains organized as a disk flow until
the magnetosphere can expand beyond $r_c$, eject inner disk
material and rebuild the magnetic tower structure of a jet. At
this point the luminosity is $L_c \sim 0.02L_{Edd}$. The
transition into the high/soft state occurs at $\sim (0.2 -
0.4)L_{Edd}$, hence there is a hysteresis of the hard
$\rightarrow$ soft transition compared to the soft $\rightarrow$
hard transition.

\section{Disk Characteristics}
For matter sufficiently inside $r_c$, the propeller mechanism is
incapable of stopping the flow, however, a boundary layer may form
at the inner disk radius in this case. The need for a boundary
layer for GBHC can be seen by comparing the magnetic pressure at
the magnetosphere with the impact pressure of a trailing, subsonic
disk. For example, for an average GBHC magnetic moment of $\sim
4\times10^{30}$ gauss cm$^3$ (see the last of Section 8), the
magnetic pressure at a $r_{ms}$ radius of $6.3\times10^6$ cm for a
7 M$_\odot$ GBHC would be $B^2 /8\pi \sim 10^{19}$ erg/cm$^3$. At
a mass flow rate of $\dot{m} = 10^{18}$ g/s, which would be near
Eddington limit conditions for a 7 M$_\odot$ MECO, the inner disk
temperature would be $T \sim 1.5\times10^7$ K. A thin disk scale
height, well behind its inner edge, would be given by $H\sim r
v_s/v_K \sim 0.0036r$, where $v_s \sim 4.5\times 10^7$cm/s and
$v_K \sim 1.2\times 10^{10}$ cm/s are acoustic and Keplerian
speeds, respectively. The impact pressure would be
$\dot{m}v_r/4\pi r H \sim 5.6\times10^5 v_r$ erg/cm$^3$. It would
require $v_r$ in excess of the speed of light to let the impact
pressure match the magnetic pressure. But since the magnetic field
doesn't eject the disk material inside $r_c$, matter piles up as
essentially dead weight against the magnetopause and pushes it in.
The radial extent of such a layer would only need to be $\sim
kT/m_pg \sim 50$ cm, where $m_p$ is the proton mass and $g$, the
radial gravitational free fall acceleration, but it is likely
distributed over a larger transition zone from co-rotation with
the magnetosphere to Keplerian flow. The gas pressure at the inner
radius of the transition zone necessarily matches the magnetic
pressure. In this case, radiation pressure in the disk, at $T=
1.5\times 10^7 K$, is nearly three orders of magnitude below the
gas pressure. Therefore a gas pressure dominated, thin, Keplerian
disk with subsonic radial speed should continue all the way to
$r_{ms}$ for a MECO. Merloni \& Fabian [2002] have shown that an
accretion disk corona can account for the hard spectrum of the low
state for a gas pressure dominated disk. Similar conditions occur
with disk radius inside $r_c$ even for weakly magnetic `atoll'
class NS. The similar magnetic pressures at $r_c$ for GBHC and
atolls is one of
the reasons for their spectral and timing similarities.

In the case of NS, sufficiently high mass accretion rates can push
the magnetopause into the star surface, but this requires near
Eddington limit conditions. At this point the hard apex of the
right side of the horizontal branch of the `Z' track in the
hardness/luminosity diagram is reached. It has recently been shown
[Muno et al. 2002] that the distinction between `atoll' and `Z'
sources is merely that this point is reached near the Eddington
limit for `Zs' and at perhaps $\sim 10 - 20$\% of this luminosity
[Barrett \& Olive 2002] for the less strongly magnetized `atolls'.
Atolls rarely reach such luminosities. For MECO based GBHC, one
would expect a relatively constant ratio of hard and soft x-ray
`colors' after the inner disk crosses $r_c$ and the flow reaches
the photon orbit. If x-ray `color' bands for GBHC were chosen
below and above a $\sim 1 keV$ thermal peak similarly to way they
are now chosen to bracket the $\sim 2 keV$ peak of NS, one might
observe
a `Z' track for the color/color diagrams of GBHC.

An observer at coordinate, r, inside $r_{ms}$, would find the
radial infall speed to be $v_r =
\frac{\sqrt{2}}{4}c(6R_g/r-1)^{3/2}$, (see Appendix A) and the
Lorentz factor for a particle spiraling in from $6R_g$ would be
$\gamma = 4 \sqrt{2}(1 + z)/3$, where $1+ z = (1-2R_g/r)^{-1/2}$
would be the red shift for photons generated at $r$. If the
distantly observed mass accretion rate would be $\dot{m}_\infty$,
then the impact pressure at r would be $p_i = (1+z)\dot{m}_\infty
\gamma v_r/ (4\pi r H)$. For $\dot{m}_\infty \sim 10^{18}$ g/s,
corresponding to Eddington limit conditions for a 7 M$_\odot$
GBHC, and $H = 0.0036r$, impact pressure is, $p_i \sim 5 \times
10^{16}(1 + z)^2(2R_g/r)^2(6R_g/r-1)^{3/2}$ erg/cm$^3$. For
comparison, the magnetic pressure is $(1 + z)^2 B_\infty^2/8\pi$.
Assuming a dipole field with average magnetic moment of $4\times
10^{30}$ gauss cm$^3$ (see Section 8), the magnetic pressure is
$\sim 10^{22} (1+z)^2 (2R_g/r)^6$ erg/cm$^3$. Thus there are no
circumstances for which the impact pressure is as large as the
magnetic pressure for $2R_g < r < 6R_g$. We conclude that another
weighty boundary layer must form inside $r_{ms}$ if the
magnetosphere is to be pushed inward. More likely, the plasma
stream is broken up by Kelvin-Helmholtz instabilities and filters
through the magnetosphere. In any event, the inner radius of the
disk is determined by the rate at which the magnetic field can
strip matter and angular momentum from the disk. This occurs in a
boundary layer of some thickness, $\delta r$, that
is only a few times the disk thickness.

Other than the presence of a transition boundary layer on the
magnetopause, the nature of the flow and spectral formation inside
$r_c$ is a research topic. Both the short radial distance from
$r_c$ to $r_{ms}$ and the magnetopause topology should help to
maintain a disk-like flow to $r_{ms}$. Radial acceleration inside
$r_{ms}$ should also help to maintain a thin flow structure. These
flows are depicted in Figure 2. Recapitulating, we expect the high
state flow into the MECO to produce a distantly observed soft
thermal component, part of which is
strongly bulk Comptonized.

{\bf Quasi-periodic Oscillations}\\
Although many mechanisms have been proposed for the high frequency
quasi-periodic oscillations (QPO) of x-ray luminosity, they often
require conditions that are incompatible with thin, viscous
Keplerian disks. Several models have requirements for lumpy flows,
elliptical inner disk boundaries, orbits out of the disk plane or
conditions that should produce little radiated power. In a
conventional thin disk, the vertical oscillation frequency, which
is approximately the same as the Keplerian frequency of the inner
viscous disk radius should generate ample power.  Accreting plasma
should periodically wind the poloidal MECO magnetic field into
toroidal configurations until the field lines break and reconnect
across the disk. Field reconnection across the disk should produce
high frequency oscillations that couple to the vertical
oscillations. If so, there would be an automatic association of
high frequency QPO with the harder power-law spectra of
magnetospherically driven emissions, as is observed. Mass ejection
in low state jets might be related to the heating of plasma via
the field breakage mechanism, in addition to natural buoyancy of a
plasma magnetic torus in a poloidal external field.

It seems possible that toroidal winding and reconnection of field
lines at the magnetopause, might continue in high states inside
$r_{ms}$. If so, there might be QPO that could be identified as
signatures of the MECO magnetosphere. If they occur deep within
the magnetosphere, they might be at locally very high frequencies,
and be observed distantly as very redshifted low frequencies. As
shown in Appendix A, the spiral orbit infall frequencies in the
plunging region inside $r_{ms}$ are given by $\nu = 1.18\times
10^5 (R_g/r)^2(1-2R_g/r)/m$ Hz. A maximum frequency of 437 Hz
would occur for m=10 at the photon orbit. Of more interest,
however are frequencies for $R_g/r \approx 1/2$, for which $\nu =
2950/(m(1+z)^2)$ Hz. For $1+z = 10 - 100, m=10$; conditions that
might apply to the photosphere region, $\nu \sim 0.03 - 3$ Hz
could be produced. In this regard, one could expect significant
time lags between inner disk accretion and luminosity fluctuations
and their echoes from the central highly
redshifted MECO.

Even if QPO are not produced inside $r_{ms}$ or inside the photon
sphere for GBHC, there is an interesting scaling mismatch that
might allow them to occur for AGN. Although the magnetic fields of
AGN scale as $m^{-1/2}$, the velocity of plasma at and inside
$r_{ms}$ does not. Thus the energy density of disk plasma inside
$r_{ms}$ will be relatively larger than magnetic field energy
densities for AGN accretion disks. When field energy density is
larger than kinetic energy density of matter, the field pushes
matter around. When the reverse is true, the matter drags the
field along. Thus toroidal winding of the field at the
magnetopause could fail to occur for GBHC, but might easily occur
in AGN. If the process is related to mass ejection, then very
energetic jets with Lorentz factors $\gamma \sim (1+z)~>10$ might
arise from within $r_{ms}$ for AGN. A field line breakage model of
`smoke ring' like mass ejection from deep within $r_{ms}$ has been
developed by Chou \& Tajima [1999]. In their calculations, a
pressure of unspecified origin was needed to stop the flow outside
$2R_g$ and a poloidal magnetic field, also of unspecified origin
was required. MECO provide the necessary ingredients in the form
of the intrinsic MECO magnetic field. The Chou \& Tajima
mechanism, aided by intense radiation pressure, may be active
inside $r_{ms}$ for GBHC and produce extremely large episodic mass
ejections such as those shown by GRS 1915+105. Although not
developed for conditions with large inner disk radius, the same
magnetic reconnection mechanism probably produces the jet
emissions [Goodson, Bohm \& Winglee 1999] associated with the
low/hard state [Gallo, Fender \& Pooley 2003].

Finally, some of the rich oscillatory behavior of GRS 1915+105 may
be readily explained by the interaction of the inner disk and the
central MECO. The objects in Table 1 have co-rotation radii of
order $(20 - 50) R_g$, which brings the low state inner disk
radius in close to the central object. A low state MECO, balanced
near co-rotation would need only a small increase of mass flow
rate to permit mass to flow on to the central MECO. This would
produce more than 20X additional luminosity and enough radiation
pressure to blow the inner disk back beyond $r_c$ and load its
mass onto the magnetic field lines where it is ejected. This also
explains the association of jet outflows with the oscillatory
states. Belloni et al. [1997] have shown that after ejection of
the inner disk of GRS 1915+105, it then refills on a viscous time
scale until the process repeats. Thus one of the most enigmatic
GBHC might be understood as a relaxation oscillator, for which the
frequency is set by a critical
mass accretion rate.

\section{Detecting MECO}
It may be possible to detect MECO in several ways. Firstly, as we
have shown, for a red shift of $z \sim 10^8$, the quiescent
luminosity of a GBHC MECO would be $\sim 10^{31}$ erg/s with
$T_\infty \sim 0.01$ keV. This thermal peak might be observable
for nearby GBHC, however it has not been found for the high
galactic latitude GBHC, XTE J1118+480. Secondly, at moderate
luminosities $L \sim 10^{36} - 10^{37}$ erg/s but in a high state
at least slightly above $L_c$, a central MECO would be a bright,
small central object that might be sharply eclipsed in deep
dipping sources. A high state MECO should stand out as a small
bright source. This is consistent with analyses of absorption dips
in GBHC GRO J1655-40 [Church 2001] which have shown the soft
source of the high state to be smaller than the region that
produces the hard spectral component of its low states. A
conclusive demonstration that most of the soft x-ray luminosity of
a high state GBHC is distributed over a large accretion disk would
be inconsistent with MECO or any other GBHC model entailing a
central bright source. If the MECO model is correct, the usual
identification of the bright, high state soft component as disk
emissions is wrong. Fitting high state spectra to multicolor disk
(MCD) blackbody models produces temperatures, $T_\infty$, that are
consistent with MECO, but due to the normalization of MCD, the
inner disk radii obtained are exactly $\sqrt{3}$ times the radius
of a MECO. The apparent constant MCD radii over a large range of
high state luminosity may merely misrepresent a constant MECO
radius that is $\sqrt{3}$ times smaller. Thirdly, a pair plasma
atmosphere in an equipartition magnetic field should be virtually
transparent to photon polarizations perpendicular to the magnetic
field lines. The x-rays from the central MECO should exhibit some
polarization that might be detectable, though this is far from
certain since the distantly observed emissions could originate
from nearly any point on the photosphere and then appear to
originate from the photon sphere. MECO presumably would not be
found only in binary systems. If they are the offspring of massive
star supernovae, then they should be found all over the galaxy. If
we have correctly estimated their quiescent temperatures, isolated
MECO-GBHC would be weak, possibly polarized, EUV sources with a
power-law tail in soft x-rays. In subsequent work we may find
additional signatures of MECO among the AGN.

\section{Conclusion}
It is now becoming apparent that many of the spectral properties
of LMXB, including the GBHC, are consistent with the existence of
intrinsically magnetized central objects. We have shown that the
existence of intrinsically magnetic GBHC is consistent with a new
class of magnetospheric eternally collapsing object (MECO)
solutions of the Einstein field equations of General Relativity.
These solutions are based on a strict adherence to the SPOE
requirement for timelike world line completeness; i.e., that the
world lines of physical matter under the influence of
gravitational and non-gravitational forces must remain timelike in
all regions of spacetime. Since there is nothing in the structure
of the Einstein tensor, $G^{\mu \nu}$, on the left hand side of
the Einstein field equation that dynamically enforces `time like
world line completeness', we have argued that the SPOE constrains
the physically acceptable choices of the energy momentum tensor,
$T^{\mu \nu}$ to contain non-gravitational forces that can
dynamically enforce it. In this context we have found the
long-lived MECO solutions. As these are necessarily based on a
radiating Vaidya metric, there is no transformation
to the Kerr-Schild coordinates used in black hole models.

An enormous body of physics scholarship developed primarily over
the last half century has been built on the assumption that
trapped surfaces leading to event horizons and curvature
singularities exist. Misner, Thorne \& Wheeler [1973], for example
in Sec. 34.6 clearly state that this is an assumption and that it
underlies the well-known singularity theorems of Hawking and
Penrose. In contrast, we have found that strict adherence to the
SPOE demand for timelike world line completeness requires a
\textit{`no trapped surface condition'}. This has led to the
quasi-stable, high red shift MECO solutions of the Einstein field
equations. The physical mechanism of their stable rate collapse is
an Eddington balance maintained by the distributed photon
generation of a a highly compact and redshifted equipartition
magnetic field. This field also serves to confine the pair plasma
dominated outer layers of the MECO and the thin MECO pair
atmosphere. Red shifts of $z \sim 10^8(m/7)^{1/2}$ have been found
to be necessary for compatibility with
our previously found magnetic moments for GBHC.

In this chapter we have given detailed descriptions of MECO
properties and shown that standard gas pressure dominated `alpha'
accretion disks would be compatible with them. We have shown that
the magnetosphere/disk interaction affects nearly all of the
spectral characteristics of NS and GBHC in LMXB systems and
accounts for them in a unified and complete way, including jet
formation and radio emissions. This model is solidly consistent
with accreting NS systems, for which intrinsic magnetic moments
obtained from spin-down measurements allow little choice. Even
their relatively weak magnetic fields are too strong to ignore.
Since the similar characteristics of GBHC are cleanly explained by
the same model, the MECO offers a unified theory of LMXB
phenomenology as well as extensions to AGN. Since MECO lifetimes
are orders of magnitude greater than a Hubble time, they provide
an elegant and unified framework for understanding the broad range
of observations associated with GBHC and AGN. Lastly we have
indicated some ways in which the existence of MECO  in GBHC and
AGN might
be detected and confirmed.\\



\section*{Appendix A. Relativistic Particle Mechanics}

A number of standard, but useful results for relativistic
mechanics are recapitulated here. All are based upon the
energy-momentum four-vector for a free particle in the
singularity-free Finkelstein or Kerr-Schild coordinates for a
constant central mass. Though not strictly compatible with
radiating objects with variable mass, outgoing Finkelstein
coordinates are a useful first order approximation to the outgoing
Vaidya coordinates for low radiation rates exterior to a MECO.
\begin{equation}
ds^2=c^2dt^2((1-2R_g/r) \pm 4R_g v^r/r - (1+2R_g/r)v^r v^r)
-r^2(d\theta^2+\sin \theta^2 d\phi^2)
\end{equation}
The plus sign corresponds to outgoing Finkelstein coordinates and
the negative sign to ingoing Finkelstein or Kerr-Schild
coordinates. Here $v^r=dr/cdt$. For a particle in an equatorial
trajectory ($\theta$ = $\pi$, $p_{\theta} = 0$) about an object of
gravitational mass M, one obtains the same equation as for
Schwarzschild coordinates:
\begin{equation}
(\frac{dr}{d\tau}) = -c(e^2-(1-2R_g/r)(1+a^2(R_g/r)^2))^{1/2}
\end{equation}
Where $e$ is the conserved energy per unit rest mass,
$a=(cp_\phi/GMm_0)$ is a dimensionless, conserved angular
momentum, $\tau$ is the proper time in the particle frame and the
negative sign indicates movement toward $r=0$. The metric Eq. (53)
also describes radial geodesics with $ds^2 = d\tau^2$. Neglecting
angular terms and letting $q=dt/d\tau$ and $p=dr/d\tau$, this
equation can be written as
\begin{equation}
1=(1-2R_g/r)q^2 \pm 4pqR_g/r - (1+2R_g/r)p^2
\end{equation}
With p given above, and $a = 0$ this equation has the solution
\begin{equation}
q= \frac{+\sqrt{e^2} \pm 2R_g/r \sqrt{e^2-(1-2R_g/r)}}{1-2R_g/r}
\end{equation}
where the positive sign on the first radical has been taken to
assure that time proceeds in a positive direction during the fall,
and a positive second term again corresponds to outgoing
coordinates. Since $v^r=p/q$, it is a straightforward matter to
substitute for $v^r$ in the original metric equation and examine
the limit as $R_g/r \rightarrow 1/2$. In outgoing coordinates, we
find that $ds^2 \rightarrow 0$ as $R_g/r \rightarrow 1/2$. 

It is of interest, however, that in the outgoing coordinates
($+4R_g/r$) as $R_g/r \rightarrow 1/2$ one finds $v^r \rightarrow
0, q \rightarrow \infty, p \rightarrow -e$. Thus it takes an
infinite coordinate time, but only a finite proper time to cross
the horizon, which is the same as the well-known Schwarzschild
result. In the ingoing coordinates, one obtains $v^r \rightarrow
0, q \rightarrow e, p \rightarrow -e$. In either case, it is
interesting to observe that the physical three-speed approaches
that of light at the horizon [Landau \& Lifshitz 1975].
\begin{equation}
V^2 = (\frac{dl}{d\tau_s})^2 = c^2\frac{({g_{0r}g_{0r}} - g_{rr}g_{00})v^r v^r}
        {(g_{00} + g_{0r} v^r)^2}
\end{equation}
Here we find $V \rightarrow c$ as $g_{00} \rightarrow 0$. Finally,
it should be mentioned that the vanishing of $g_{00}$ for $r > 0$
is actually a result of a failure to apply appropriate boundary
conditions for the solutions of the Einstein equations for a point
mass
[Abrams 1979, 1989].

For suitably small energy, bound orbits occur. Turning points for
which $dr/d\tau = 0$ can be found by examining the effective
potential, which consists of all terms to the right of $e^2$ in
Eq. (54). At minima of the effective potential we find circular
orbits for which
\begin{equation}
a^{2}=\frac{1}{R_g/r-3 (R_g/r)^{2}}
\end{equation}
$R_g/r=1/3$ holds at the location of an
unstable circular orbit for photons (see below).
From which we see that if $p_\phi$ is non-zero there are no trajectories
for particles with both mass and angular momentum that exit from within $R_g/r=1/3$.
Thus particles with both mass and angular momentum
can't escape from within the photon sphere. The minimum energy
required for a circular orbit would be.
\begin{equation}
E=m_{0}c^{2}\frac{(1-2 R_g/r)}{\sqrt{(1-3 R_g/r)}}
\end{equation}
In fact, however, there is an innermost marginally stable orbit for
which the first two derivatives with respect to 1/r of the
effective potential vanish. This has no Newtonian physics counterpart,
and yields the well-known results: $R_g/r=1/6$, $a^2 = 12$ and $e^2 =8/9$
for the marginally stable orbit of radius $r_{ms} = 6GM/c^2$.

For a particle beginning a spiral descent from $r_{ms}$ with $e=\sqrt{8/9}$,
there follows:
\begin{equation}
(\frac{dr}{d\tau})^2=c^2\frac{(6R_g/r-1)^3}{9}
\end{equation}
If observed by a stationary observer located at coordinate r, it
would be observed to move with radial speed
\begin{equation}
V_r = \frac{\sqrt{2}c(6R_g/r-1)^{3/2}}{4}.
\end{equation}
Again, $V_r$ approaches c as $R_g/r$ approaches 1/2.
A distant observer would would find the angular frequency of the spiral motion
to be

\[
\frac{1}{2\pi}\frac{d\phi}{dt} = \sqrt{9\times 12/8}(c^3/GM)(R_g/r)^2(1-2R_g/r)/2\pi
\]
\begin{equation}
\sim 1.18\times 10^5 (R_g/r)^2(1-2R_g/r)/m ~~~Hz
\end{equation}
For a 10 M$_\odot$ GBHC ($m = 10$), this has a maximum of 437 Hz and
some interesting possibilities for generating many QPO frequencies, both
high and low. For red shifts such that $R_g/r \approx 1/2$, the spiral
frequency is  $2950/(1+z)^2$ Hz.

\subsection*{Photon Trajectories:}
 The energy-momentum
equation for a particle with $m_0=0$ can be rearranged as:
\begin{equation}
(1-2R_g/r)^2(\frac{p_rGM}{p_\phi c^2})^2 = (\frac{d(R_g/r)}{d\phi})^2 =
(\frac{GME}{p_\phi c^3})^2 - (R_g/r)^2(1-2R_g/r)
\end{equation}
The right member has a maximum value of 1/27 for $R_g/r=1/3$. There is an
unstable orbit with $d(R_g/r)/d\phi = 0$ for $R_g/r=1/3$. To simply have
$d(R_g/r)/d\phi$ be real requires $p_\phi c^3/GME < \sqrt{27}$. But
$E=(1+z)p c$, where p is the entire momentum of the photon, and
$1+z = (1-2R_g/r)^{-1/2}$ its red
shift if it escapes to be observed at a large distance. Its azimuthal
momentum component will be $p_\phi/r$. Thus its escape cone is defined by:
\begin{equation}
(\frac{p_\phi}{r p})^2 < 27(R_g/r)^2(1-2R_g/r)
\end{equation}\\
For $r_s \approx 2R_g$, this approaches $27/(4(1+z_s)^2)$.

\section*{Appendix B. Pair Plasma Photosphere Conditions}

The photosphere condition is that [Kippenhahn \& Weigert 1990):
\begin{equation}
n_\pm \sigma_T l = 2/3,
\end{equation}
where $n_\pm$ is the combined number density of electrons and
positrons in equilibrium with a photon gas at temperature T,
$\sigma_T = 6.65\times 10^{-25} cm^2$ is the Thompson scattering
cross section and $l$ is a proper length over which the pair
plasma makes the transition from opaque to transparent. Landau \&
Lifshitz [1958] show that
\begin{equation}
n_\pm=\frac{8\pi}{h^3}\int_{0}^{\infty}
\frac{p^2dp}{\exp{(E/kT)}+1}
\end{equation}
where p is the momentum of a particle, $E=\sqrt{p^2c^2+m_e^2c^4}$,
k is Boltzmann's constant, $h$ is Planck's constant and $m_e$, the mass of an
electron. For low temperatures such that $kT < m_ec^2$ this becomes:
\begin{equation}
n_\pm \approx 2(\frac{2 \pi m_ekT}{h^2})^{3/2} \exp{(-m_ec^2/kT)}
=2.25\times 10^{30}(T_9/6)^{3/2} \exp{(-6/T_9)}~/cm^3
\end{equation}
where $T_9=T/10^9$K.
\bigskip

For a photosphere temperature of $4\times 10^8$K, for $m=7$, the
number density of pairs is $n_\pm=1.4\times 10^{22}~/cm^3$, with
the proper density reduced from this value by $1+z_p$. The mean
free path of photons among these is $\sim 100~cm$ for a Thompson
scattering cross-section of $6.65\times 10^{-25}~cm^2$.
Considering the extreme redshift variation over a small radial
coordinate interval, (Eq. (27)), the proper interval corresponding
to a mean free path of $\sim 100~cm$ would be expanded by a
redshift factor somewhere between $1+z_p$ and $(1+z_s)$ to the
range $10^{5, 10}~cm$; i.e, to a normal and reasonable photosphere
depth. To a distant observer, the pair atmosphere would occur in
an extremely small coordinate interval. Modeling it numerically
would be very difficult. As previously noted. the photosphere
temperature is very nearly independent of the mass of the MECO. 

At a surface temperature of $6\times 10^9$ K, we find a proper
pair density of $n_\pm \sim 2.25\times 10^{30}/(1+z_s)~/cm^3$,
which produces a maximum possible pair pressure $\sim n_\pm m_e
c^2/3$ some $10^9$ times less than the radiation pressure. This
justifies our use of radiation dominated pressure in the pair
atmosphere. As shown in Appendix D, we expect $n_\pm \sim
10^{22}~/cm^3$ from consideration of the surface magnetic fields.
For a $m=10$ MECO, this is just about what we get for $T=6\times
10^9$ K.


\section*{Appendix C. New Observations}

The third accreting millisecond pulsar, {\bf XTE J0929-314} has
been found [Galloway et al. 2002] with $\nu_s = 1/P = 185$ Hz and
period derivative $\dot{P} = 2.69\times 10^{-18}$, from which the
magnetic field (calculated as $3.2\times \sqrt(P\dot{P})$ is
$3.9\times 10^9$ gauss. This is typical of a Z source. Assuming a
NS radius of 13 km, the magnetic moment is $BR^3=8.5\times
10^{27}$ gauss cm$^3$. The calculated low state limit co-rotation
luminosity is $L_c=4.9\times 10^{36}$ erg/s. Approximately 40\% of
this would be the luminosity in the (2 - 10 keV) band. This yields
an expected flux of $2\times 10^{-10}$ erg/cm$^2$/s for a distance
of 9 kpc. This corresponds to the knee of the published light
curve where the luminosity begins a rapid decline as the propeller
becomes active. Similar breaking behavior has been seen in Sax
J1808.4-3659 and GRO J1655-40 at propeller onset. The predicted
0.5-10 keV band luminosity
is $L_q = 1.3\times 10^{33}$ erg/s.

The second accreting millisecond pulsar {\bf XTE J1751-305} was
found with a spin of 435 Hz. [Markwardt et al. 2002] Its spectrum
has been analyzed [Miller et al. 2003]. We find a hard state
luminosity of $3.5\times 10^{36}$ erg/s ($d=8$ kpc) at the start
of the rapid decline which is characteristic of the onset of the
propeller effect. We take this as an estimate of $L_c$. From this
we estimate a magnetic moment of $1.9\times 10^{27}$ gauss cm$^3$
and a quiescent luminosity of $5\times 10^{33}$ erg/s. An upper
limit on quiescent luminonosity of $1.8\times 10^{34}$ erg/s can
be set by the detections of the source in late April 2002, as
reported by Markwardt et al. [2002].

The accreting x-ray pulsar, {\bf GRO J1744-28} has long been cited
for exhibiting a propeller effect. Cui [1997] has given its spin
frequency as 2.14 Hz and a low state limit luminosity as $L_c =
1.8\times 10^{37}$ erg/s (2 - 60 keV.), for a distance of 8 kpc.
These imply a magnetic moment of $1.3\times 10^{31}$ gauss cm$^3$
and a magnetic field of $B=5.9\times 10^{12}$ gauss for a 13 km
radius. It spin-down energy loss rate should be $\dot{E} =
1.4\times 10^{35}$ erg/s and its quiescent luminosity, $L_q =
3\times 10^{31}$ erg/s. Due to its slow spin, GRO J1744-28 has a
large co-rotation radius of 280 km. A mass accretion rate of
$\dot{m}= 5.4\times 10^{18}$ g/s is needed to reach $L_c$. Larger
accretion rates are needed to reach the star surface, but such
rates distributed over the surface would produce luminosity in
excess of the Eddington limit. The fact that the magnetic field is
strong enough to funnel a super-Eddington flow to the poles is the
likely reason for the type II bursting behavior sometimes seen for
this source. In addition to its historical illustration of a
propeller effect, this source exemplifies the inverse correlation
of spin and magnetic field strength in accreting sources. It
requires a weak field to let an accretion disk get close enough to
spin up the central object. For this reason we expect Z sources
with their stronger B fields to generally spin more
slowly than atolls.

The accreting pulsar, {\bf 4U0115+63}, with a spin of 0.276 Hz and
a magnetic field, derived from its period derivative, of
$1.3\times 10^{12}$ gauss (yielding $\mu = 2.9\times 10^{30}$
gauss cm$^3$ for a 13 km radius) has been shown [Campana et al.
2002] to exhibit a magnetic propeller effect with a huge
luminosity interval from $L_c = 1.8\times 10^{33}$ erg/s to
$L_{min} = 9.6\times 10^{35}$ erg/s. $L_c$ held steady precisely
at the calculated level for a lengthy period before luminosity
began increasing. Due to the slow spin of this star, its quiescent
luminosity, if ever observed, will be just that emanating from the
surface. Its spin-down
luminosity will be much too low to be observed.

The atoll source {\bf 4U1705-44} has been the subject of a recent
study [Barret \& Olive 2002] in which a Z track has been displayed
in a color-color diagram. Observations labelled as 01 and 06 mark
the end points of a spectral state transition for which the
luminosity ratio $L_{min}/L_c = 25.6\times 10^{36}/6.9\times
10^{36} = 3.7$ can be found from their Table 2. These yield $\nu =
470$ Hz and a magnetic moment of $\mu = 2.5\times 10^{27}$ gauss
cm$^3$. The spin-down energy loss rate is $1.2\times 10^{37}$
erg/s and the 0.5 - 10 keV quiescent luminosity is estimated to be
about $5\times 10^{33}$ erg/s. At the apex of the Z track
(observation 12), the luminosity was $2.4\times 10^{37}$ erg/s
(for a distance of 7.4 kpc.); i.e., essentially the same as
$L_{min}$. Although 4U1705-44 has long been classified as an atoll
source, it is not surprising that it displayed the Z track in this
outburst as
its 0.1 - 200 keV luminosity reached 50\% of the Eddington limit.

Considerable attention was paid to reports of a truncated
accretion disk for the GBHC, {\bf XTE J1118+480} [McClintock et al
2001] because of the extreme interest in advective accretion flow
(ADAF) models for GBHC [Narayan, Garcia \& McClintock 1997].
McClintock et al, fit the low state spectrum to a disk blackbody
plus power law model and found that the disk inner radius would be
about 35R$_{schw}$, or 720 km for 7 M$_\odot$. Using this as an
estimate of the co-rotation radius we find the spin to be 8 Hz.
The corresponding low state luminosity of $1.2\times 10^{36}$
erg/s (for $d= 1.8$ kpc) lets us find a magnetic moment of
$10^{30}$ gauss cm$^3$. The calculated spin-down energy loss rate
is $1.5\times 10^{35}$ erg/s and the quiescent luminosity
would be about $3\times 10^{31}$ erg/s.

A rare transition to the hard state for {\bf LMC X-3} [Soria, Page
\& Wu 2002, Boyd et al. 2000] yields an estimate of the mean low
state luminosity of $L_c = 7\times 10^{36}$ erg/s and the high
state luminosity in the same 2 - 10 keV band is approximately
$6\times 10^{38}$ erg/s at the end of the transition to the soft
state. Taking these as $L_c$ and $L_{min}$ permits the estimates
of spin $\nu = 16$ Hz and magnetic moment $\mu = 8.6\times
10^{29}$ gauss cm$^3$, assuming 7 M$_\odot$. From these we
calculate a quiescent luminosity of $10^{33}$ erg/s.\\
%

\section*{Appendix D. The Existence and Stability of Highly Redshifted MECO}
A MECO is, in many ways, more exotic than a black hole with its
mere mass and spin. It is equally compact but its surface magnetic
field is sufficient to produce bound pairs from the quantum
electrodynamic vacuum. This occurs at a threshold that is
insensitive to mass, thus MECO can range in mass from the GBHC to
AGN. As we shall see, the scaling with mass of the distantly
observed magnetic fields, $B \propto M^{-1/2}$ allows the ratio
$L_c/L_{Edd}$ to also be mass scale invariant [Robertson \& Leiter
2004]. The MECO interior magnetic fields are relatively modest.
Interior and surface fields differ due to substantial pair drift
currents on the MECO surface. In general, plasmas in hydrostatic
equilibrium in magnetic and gravitation fields experience drift
currents proportional to ${\bf g\times B}/B^2$. The general
relativistic generalization of this
provides the key to our understanding of the high redshifts of MECO.

The surface temperature and high luminosity to radius ratio
(hereafter L/R compactness, see appendix F). of the MECO Eddington
limited, timelike, secular collapsing state implies that the
plasma is dominated by electron-positron pairs. These are
generated by colliding photons due to the optically thick
synchrotron luminosity of the intrinsic MECO magnetic field, both
within the interior and on the MECO surface. Recall that the
surface is well inside the photon orbit and the bulk of the photon
outflow from the surface falls back. The existence of the MECO
state requires that:
\begin{equation}
L_{Edd}(outflow)\sim L_{Syn}(out)
\end{equation}
within the MECO and
\begin{equation}
L_{Edd,S}(escape) \sim L_{Syn,S}(escape)
\end{equation}
at the MECO surface S. Where (see Section 5, Eq. (22))
\begin{equation}
L_{Edd,S}(escape) \sim (4\pi G M_s(\tau)c /\kappa)(1 + z_s) \sim
1.3\times10^{38}m(1+z_s)
\end{equation}
\begin{equation}
L_{Syn,S}(escape) \sim L_{Syn,S}(out)/{(4/27)(1+z_s)^2}
\end{equation}

Assuming a temperature near the pair production threshold, the
rate of synchrotron photon energy generation in a plasma
containing $N_\pm$ electrons and positrons is [Shapiro \&
Teukolsky 1983]
\begin{equation} L_{Sync,S}(out) \sim
(16e^4/3m_e^2c^3)N_\pm B^2(T_9/6)^2 \sim 1.27 \times 10^{-14}N_\pm
B^2(T_9/6)^2~~erg/s
\end{equation}
where $T_9=10^-9T$ and $T_9/6=kT/m_ec^2$.

For an Eddington equilibrium, we require the synchroton generation
rate to produce the outflow through the MECO surface. Thus
\begin{equation}
L_{Edd}(out) \sim 1.27 \times 10^{38} m (4(1+z_s)^3/27) \sim 1.27
\times 10^{-14}N_\pm B^2(T_9/6)^2~~erg/s
\end{equation}
which implies that
\begin{equation}
N_\pm B^2 \sim 10^{52}(m/7)(1  + z_s)^3(6/T_9)^2~~erg/cm^3
\end{equation}
\bigskip

From Section 6 of Baumgarte \& Shapiro [2003], we note that if
$\mu$ is the distantly observed MECO magnetic moment and
$(1+z_s)>>1$ is the MECO surface redshift, then the Einstein-
Maxwell equations imply that the components of the MECO dipole
magnetic field strength, B at distance r are given by
\begin{equation}
B_r=2F(x)\mu \cos(\theta)/r^3
\end{equation}
and
\begin{equation}
B_{\theta}=G(x)\mu \sin(\theta)/r^3
\end{equation}
where $x=r_s/2R_g$ and
\begin{equation}
F(x)=(-3x^3)(ln(1-x^{-1})+x^{-1}(1+x^{-1}/2))
\end{equation}
\begin{equation}
G(x)=(6x^3)((1-x^{-1})^{1/2}ln(1-x^{-1})+x^{-1}(1-x^{-1}/2)(1-x^{-1})^{-1/2})
\end{equation}
Note that for $r_s>>2R_g$, $x>>1$ and both $F(x)$ and $G(x)
\rightarrow 1$, while as we approach a compact MECO surface where
$(1+z_s) >>> 1$, then $x \rightarrow 1^+$ and
\begin{equation}
F(x) \rightarrow -3ln(1-x^{-1}) = -3ln(1/(1+z_s)^2) = 6ln(1+z_s)
\end{equation}
and
\begin{equation}
G(x)=3(1-x^{-1})^{-1/2}=3(1+z_s)
\end{equation}
Hence the radial component of the magnetic field on the MECO
surface is given by
\begin{equation}
B_{r,S^+} =12ln(1+z_s)\mu \cos(\theta)/(2R_g)^3
\end{equation}
while the poloidal component is given by
\begin{equation}
B_{\theta,S^+}=3\mu(1+z_s)\sin(\theta)/(2R_g)^3
\end{equation}
The interior magnetic dipole fields, $B'$ in the MECO, which are
due to the interior MECO magnetic dipole moment $\mu(r)$ in the
interior will be given by
\begin{equation}
B'_r=12\mu(r) \cos(\theta)\ln(1+z)/r^3
\end{equation}
and
\begin{equation}
B'_{\theta}=6\mu(r) \ln(1+z) \sin(\theta)/r^3
\end{equation}
Thus the expressions for the exterior magnetic field just outside
of the MECO surface differ via $F(x_s)$ and $G(x_s)$ from those of
the interior magnetic field. But these
expressions have the following important consequences:

{\indent \it (a) The general relativistic structure of the
Maxwell-Einstein equations causes the radial and poloidal exterior
components of the MECO magnetic dipole fields to undergo redshift
effects which are different functions of $(1 + z_s)$\\
(b) The radial component $B_{r,S}$ of the magnetic dipole field is
continuous at the MECO surface, but the poloidal component is not.
$B_{\theta,S^+}$ is different from $B'_{\theta,S^-}$ at the MECO
surface.}

This difference is caused by powerful $e^{\pm}$ drift currents
that are induced by the strong gravitational field and enhanced by
the differing general relativistic dependence on redshift of the
poloidal and radial magnetic field components. This is a general
relativistic generalization of the fact that a plasma in
hydrostatic equilibrium in gravitational and magnetic fields
experiences drift currents proportional to ${\bf g\times B}/B^2$.
In fact, it is the drift currents that generate the
distantly observed magnetic moments seen in MECO-GBHC and MECO-AGN.

The MECO magnetic moment coupling to a surrounding accretion disk
will cause it to be a slow rotator. Hence to estimate the strength
of the magnetic field just under the surface of the MECO of the
GBHC we can use the results obtained for low redshift, slowly
rotating compact stellar objects with magnetic fields. The
magnetic field strength in the interior of a slowly rotating
neutron star of radius $\sim 10$ km, was shown to be $\sim
10^{13}$G [Gupta, Mishra, Mishra \& Prasanna 1998]. When scaled to
the $\sim 7 M_\odot$ and $R \sim 2R_g$ size of the MECO, the
magnetic fields under the surface can be estimated to be
(neglecting latitude angle dependence):
\begin{equation}
B_{r,S^-} \sim (2\mu/(2R_g)^3)6 ln(1+z_s) \sim (10^{13.7} gauss)/
(M / 7M_\odot)^{1/2}
\end{equation}
\begin{equation}
B_{\theta,S^-} \sim (\mu/(2R_g)^3)6 ln (1+z_s) \sim (10^{13.4}
gauss)/ (M / 7M_\odot)^{1/2}
\end{equation}
Then from Eq.s (81), (82) and (85), the exterior magnetic fields
on the MECO surface S are:
\begin{equation}
B_{r,S^+} = B_{r,S^-} \sim (2\mu/(2R_g)^3)6 ln(1+z_s)cos(\theta)
\sim (10^{13.7} gauss) cos(\theta) /(M/7M_\odot)^{1/2}
\end{equation}
\begin{equation}
B_{\theta,S^+} \sim (\mu / (2R_g)^3)3 (1+z_s) sin(\theta)
\end{equation}
Using these equations,  the magnitude of the surface redshift
$(1+z_s)$  for a MECO can be directly determined by requiring that
the strength of the poloidal component of the equipartition
magnetic field $B_{\theta,S}$ on the MECO surface:  (a)  must be
mass scale invariant and,  (b) cannot be much larger the quantum
electrodynamically determined maximum value for a NS given by
$B_{\theta,S} \sim 10^{20}$ gauss [Harding, A., 2003]. This is
because surface magnetic fields much larger than $\sim 10^{20}$
gauss would create a spontaneous quantum electrodynamic phase
transition associated with the vacuum production of bound pairs on
the MECO surface [Zaumen 1976]. This would cause more pairs to be
produced than those required by the Eddington balance of the
MECO-GBHC surface. This would then cause the MECO-GBHC surface to
expand. However the resultant expansion due to this process would
reduce the redshift and the surface poloidal magnetic field thus
quenching the vacuum production of bound pairs and allowing the
MECO-GBHC surface to contract. This stability mechanism on the
MECO surface implies that its surface redshift $(1+z_s)$ can be
dynamically determined from the preceding pair of equations.
Neglecting the trigonometric functions common to both sides of the
equations, the ratio of these external field components in Eq.s
(88) and (86), yields
\begin{equation}
3(1+z_s)/[6ln(1+z_s)] \sim 10^{20}/[10^{13.4}/(m/ 7)^{1/2}]
\end{equation}
for which the solution is
\begin{equation}
(1+z_s) \sim 1.5\times 10^8(m/7)^{1/2}
\end{equation}
In addition we obtain
\begin{equation}
\mu / (2R_g)^3 \sim (2.2\times 10^{11} gauss)/(m / 7)^{1/2}
\end{equation}
{\it which implies that the average distantly observed intrinsic
magnetic moment of the MECO is
\begin{equation}
\mu \sim (2\times 10^{30}gauss~ cm^3)(m / 7)^{5/2}
\end{equation}
This is in good agreement with our analysis of observations. (See
Table 1 and comments near the end of Section 8.)}

Using Eq. (74) we can now get a rough estimate of the pair
density, by considering the $N_\pm$ to be uniformly distributed
over a volume of $4(1+z_s) \pi (2R_g)^3/3$ and by considering the
interior magnetic field to be uniform at $2.5\times
10^{13}/(m/7)^{1/2}~gauss$. Hence
\begin{equation}
n_\pm=10^{22} (m/7)^{-3/2}(6/T_9)^2~~/cm^3
\end{equation}
\bigskip
which for a GBHC, agrees within a factor of two of the result
found for a pair plasma at $T=6\times 10^9$K. This result for
$n_\pm$ implies that surface temperature would increase with
increasing mass, however, it only increases by a factor of $10$
for $m=10^8$. Since mean MECO densities scale as $1/m^2$, one
might expect larger density gradients and different ratios of
pairs to neutrons and protons in AGN compared to GBHC, which are
approximately
of nuclear densities.

Azimuthal MECO surface currents  are the source of the distantly
observed magnetic moment seen in the MECO-GBHC. The magnitude of
these surface currents is essentially mass scale invariant and is
given by
\begin{equation}
i(S) = (c / 4\pi )(\mu /(2R_g)^3)[3(1+z_s)]sin(\theta) \sim
3.3\times 10^{26}~~amp/cm^2
\end{equation}
total current on GBHC surface, which corresponds to
\begin{equation} \sim 2\times 10^{45} e^\pm /sec
\end{equation}
combined surface e(+ -) flow. Hence (94) and (95) imply that the
corresponding drift speeds of electrons and positrons are $v/c
\sim 1$. This implies that the opposed e(+-) pairs currents on the
MECO surface are moving relativistically and hence will have a
very long lifetime before annihilating. This maintains a stable
flow of current as required to generate the distantly observed
MECO magnetic moments.

The radiation pressure at the outer surface of the MECO is
\begin{equation}
P_{Syn}(out)= L_{Syn}(out)/[4\pi(2R_g)^2c]=
 \sim 2.7\times 10^{38}(m/7)^{1/2}~~erg/cm^3
\end{equation}
For comparison, the mass-energy density of a MECO is
\begin{equation}
\rho c^2 \sim Mc^2/[(1+z_s)4\pi (2R_g)^3/3] \sim 2\times
10^{27}/(m/7)^{2.5}
\end{equation}
\bigskip
which suggests that MECO is radiation dominated and very tightly
gravitationally bound. An order of magnitude calculation of
binding energy \footnote{This has obvious, important consequences
for hypernova models of gamma-ray bursters.} yields $\sim 1.5 Mc^2
~ln(1+z_s)$ for a residual mass, M. Thus the progenitor of a a
MECO-GBHC would have a mass of 200-300 $M_\odot$, and this
suggests that they would be among the earliest and most massive
stars in the galaxy. A $10^9 M_\odot$ MECO-AGN could originally
consist of $4\times 10^{10}M_\odot$.\\
%

\section*{Appendix E.  The $\gamma + \gamma  \leftrightarrow e^\pm$   Phase Transition
and MECO Existence}
It is well-known that a spherical volume of radius $R$  containing
a luminosity $L_{\gamma}$  of  gamma ray photons with energies $>$
1 MeV,\footnote{1 MeV photons correspond to $T \sim 10^{10}~K$,
which is only slightly beyond the pair threshold, and easily
within reach in gravitational collapse.} will become optically
thick to the $\gamma + \gamma \leftrightarrow e^\pm$ process when
\begin{equation}
\tau_\pm \sim n_\gamma\sigma_{\gamma\gamma}R \sim 1
\end{equation}
and
\begin{equation}
n_\gamma \sim L_\gamma/(2\pi R^2 m_ec^3)
\end{equation}
is the number density of $\gamma$-ray photons with energies $\sim$
1 MeV, $L_\gamma$ is the $gamma$-ray luminosity, $R$ is the radius
of the volume, and $\sigma{\gamma\gamma}$ is the pair production
cross section.

Since $\sigma{\gamma\gamma} \sim \sigma_T$ near threshold, it
follows that the system becomes optically thick to  photon-photon
pair production when the numerical value of its compactness
parameter $L_\gamma/R$ is
\begin{equation}
L_\gamma/R \sim 4\pi m_e c^3/\sigma_T \sim 5\times 10^{29}~~~
erg/cm-sec
\end{equation}
Hence $\tau_\pm > 1$ will be satisfied for systems with
compactness
\begin{equation}
L_\gamma/R > 10^{30}~~erg/cm-sec
\end{equation}
For an Eddington limited MECO, which has a very large surface
redshift $(1 + z)  >> 1$  at  $R \sim 2R_g$ , and taking the
proper length and volume into consideration, the optical depth to
photon-photon pair production has the very large value
\begin{equation}
\tau_\pm(1+z) \sim (L_{\gamma, 30}/R) \sim 10^2 \times (1+z)^3
>> 1~~erg/cm-sec
\end{equation}
Thus the resultant $\gamma + \gamma \leftrightarrow e^\pm$ phase
transition in the MECO magnetic field, $B_S$, creates an optically
thick pair dominated plasma. Taking the photon escape cone factor
$\sim  1 /(1+z)^2$  into account, the process generates a net
outward non-polytropic radiation pressure (see Appendix D)
\begin{equation}
P \sim (1 + z) B_S^4 m
\end{equation}
on the MECO surface. The increase of pressure with redshift is a
key feature of the Eddington limited secular balance at $R \sim
2GM/c^2$. Thus trapped surfaces, which lead to event horizons, can
be prevented from forming. As discussed in Appendix D., the
balance is mass scale invariantly stabilized at the threshold of
magnetically produced pair breakdown of the vacuum.\\

\section*{Appendix F - On The Black Hole Kerr-Schild Metric And
MECO Vaidya Metric Solutions To the Gravitational Collapse
Problem}

In discussions with experts in general relativity the validity of
our motivation to look for physical alternatives to black holes
has been questioned. Our work has been based on the assumption
that the preservation of the Strong Principle of Equivalence
(SPOE) in Nature implies that metrics with event horizons are
non-physical. The objection to this has been based on the well
known fact that for massive particles under the action of both
gravitational and non-gravitational forces, the timelike nature of
the world line of massive particles is preserved. The generally
covariant equation of motion for their timelike world lines in
spacetime is given by
\begin{equation}
Du^{\mu} / d\tau = a^{\mu} = K^{\mu}
\end{equation}
Here $u^{\mu}$ is the four velocity of the massive particle and
$K^{\mu}$ is the generally covariant non-gravitational four-vector
force which in general relativity is required to obey the dynamic
condition
\begin{equation}
K^{\mu}a_{\mu} = 0
\end{equation}
Then from the above two equations it follows that
\begin{equation}
D(u^{\mu}u_{\mu}) / d\tau = 0
\end{equation}
which guarantees that [where we have chosen units where c=1 and
use the spacetime metric signature (1,-1,-1,-1)]
\begin{equation}
u^{\mu}u_{\mu} =1
\end{equation}
From this it follows that in general relativity the timelike
invariance of the world line of a massive particle is dynamically
preserved for all metric solutions, $g_{\mu\nu}$, to the Einstein
Equations, including the case of the ``event horizon penetrating"
Kerr-Schild metric used by most black hole theoreticians in
computer simulations of the black hole collapse of a radially
infalling massive particle or fluid. On the basis of the above
facts it is then argued that there is no reason to look for
physical alternatives to black holes and that the assumption that
the preservation of the Strong Principle of Equivalence (SPOE) in
Nature implies that metrics with event horizons are non-physical,
is in error.

However we will now show that above arguments, which are based on
the four velocity $u^{\mu}$ alone, are not valid. This is because
relativists who come to this conclusion in this manner are making
the mistake of ignoring the fact that in addition to the four
velocity $u^{\mu}$ there exists another important quantity called
the ``physical 3-velocity" which must also be considered as well.
Physically speaking, the magnitude of the physical 3-velocity is
seen by an observer at rest as being equal to the speed of the
co-moving observer who is moving along with the collapsing massive
particle or fluid. If we consider the case of a radially infalling
massive particle or fluid undergoing gravitational collapse it can
be shown that the radial component of the physical 3-velocity is
given by
\begin{equation}
V^r = c \frac{[(g_{0r}g_{0r} - g_{rr}g_{00})v^r
v^r]^{1/2}}{|(g_{00}+ g_{0r}v^r)|}
\end{equation}
where $v^r = dr/d\tau$ is the radial coordinate velocity of the
massive particle of fluid (See Landau and Lifshitz., 1975
``classical Theory of Fields", 4th Ed, Pergamon Press pg 248-252).

From the above formula for the radial component of the physical
3-velocity of the co-moving observer $V^r$ we see that for metrics
which have the property that $g_{00} \rightarrow 0$ in some region
of spacetime (i.e. the property associated with the existence of
an event horizon for the non-rotating metrics associated with the
radial infall of matter) the physical radial velocity $V^r$ of the
co-moving frame of the massive particle or fluid becomes equal to
the speed of light as the massive particle or fluid crosses the
event horizon.

Hence even though the timelike property $u^{\mu} u_{\mu} = 1$ is
preserved for a massive particle or fluid crossing the event
horizon of the Kerr-Schild metric where $g_{00} \rightarrow 0$
occurs, a local special relativistic connection between the
co-moving frame of the radially infalling massive particle or
fluid and a stationary observer can no longer be made. Since the
Strong Principle Of Equivalence (SPOE) requires that Special
Relativity must hold locally at all points in spacetime, the
breakdown at the event horizon, of the local special relativistic
connection between the co-moving observer frame and a stationary
observer frame for a particle crossing the event horizon,
represents a violation of the SPOE. Hence we have shown that by
considering both the four velocity and the ``physical 3-velocity
of the co-moving observer" there is motivation to look for
physical alternatives to black holes. In fact in this context
logical arguments can be consistently made which show that Black
Holes with non-zero mass cannot exist in Nature (Mitra, A., 2000,
2002, 2005, 2006).

Based on the above arguments, the SPOE preserving requirement that
the co-moving observer frame for a massive collapsing fluid must
always be able to be connected to a stationary observer by special
relativistic transformations with a physical 3-speed which is less
than the speed of light, was taken seriously in our work. In the
literature the requirement that the SPOE must be preserved
everywhere in spacetime for the timelike worldlines of massive
particles or fluids under the influence of both gravitational and
non-gravitational forces goes under the technical name of
``timelike worldline completeness".

Based on this idea we have found that preservation of the SPOE in
Nature can be accomplished only if there exist non-gravitational
components in the energy-momentum tensor on the right hand side of
the Einstein equation that physically guarantee the preservation
of the SPOE. It was in this alternative context that the general
relativistic MECO solutions to the Einstein-Maxwell equations
emerged, as was shown in the three previously published papers of
Robertson and Leiter and developed in more detail in Appendix 1-10
in this paper. There it was shown that for a collapsing body, the
structure and radiation transfer properties of the energy-momentum
tensor on the right hand side of the Einstein field equations,
could describe a collapsing radiating object which contained
equipartition magnetic fields that generated a highly redshifted
Eddington limited secular collapse process. This collapse process
was shown to satisfy the SPOE requirement of Timelike Worldline
Completeness by dynamically preventing trapped surfaces, that lead
to event horizons, from forming.

More specifically in Appendix A-E it was shown that, by using the
Einstein-Maxwell Equations and Quantum Electrodynamics in the
context of General Relativistic plasma astrophysics, it was
possible to virtually stop and maintain a slow, (many Hubble
times!) steady collapse of a compact physical plasma object
outside of its Schwarzschild radius. The non-gravitational force
was Compton photon pressure generated by synchrotron radiation
from an intrinsic equipartition magnetic dipole field contained
within the compact object. The rate of collapse is controlled by
radiation at the local Eddington limit, but from a highly
redshifted surface. In Appendix D it was shown that general
relativistic surface drift currents within a pair plasma at the
MECO surface can generate the required magnetic fields. In
Appendix D the equatorial poloidal magnetic field associated with
a locally Eddington limited secular rate of collapse of the
exterior surface was shown to be strong enough to spontaneously
create bound electron-positron pairs in the surface plasma of the
MECO. In the context of the MECO highly redshifted Eddington
limited balance, the action of this QED process was shown to be
sufficient to stabilize the collapse rate of the MECO surface.

For the case of hot collapsing radiating matter associated with
the MECO, the corresponding exterior solution to the Einstein
equation was shown to be described by the time dependent Vaidya
metric. No coordinate transformation between MECO Vaidya metric
and the Black Hole Kerr-Schild metric exists. Since the highly
redshifted MECO Vaidya metric solutions preserve the SPOE and they
do not have event horizons, they can also contain a slowly
rotating intrinsic magnetic dipole moment. These magnetic moments
have observable effects if such MECO exist at the centers of
galactic black hole candidates and AGN. In recent work (Schild,
Robertson \& Leiter 2005) we have found observational evidence of
the physical effects of such intrinsic magnetic dipole fields in
the central compact object in the Quasar Q0957+561. It is
important to note that these physical effects cannot be explained
in terms of standard Black Hole models using Kerr-Schild metric
driven GRMHD calculations, since these calculations generate
unphysical ``split magnetic monopole fields" that cannot explain
the details of the intrinsic structure in Q0957+561 that our
observations have found.

We reiterate that the Kerr-Schild metric used by most relativists
is not relevant to the work done in this paper, because the
collapsing radiating MECO solution to the Einstein equation is
described by the time dependent radiating Vaidya metric, and there
is no coordinate transformation between them. Therefore we found
that we had to turn to alternate MECO Vaidya metric solutions to
the Einstein-Maxwell equations which feature the intrinsic
magnetic dipole fields implied by the observations.

\subsection*{Acknowledgements}
We thank Abhas Mitra for seminal ideas, helpful comments and
criticisms and many discussions of issues related to gravitational
collapse. Useful information has been generously provided by Mike
Church, Heino Falcke,  Elena Gallo, Thomas Maccarone and Jeff
McClintock.

\section*{References}

\newenvironment{refhanging}{
\vspace{-0.35in}
\begin{list}{}{
  \topsep          5mm
  \leftmargin      6mm
  \parsep          1mm
  \listparindent     -0.25in}
  \item[]\strut\par}{
\end{list}}

\begin{refhanging}

\item [Abramowicz, M., Kluzniak, W. \& Lasota], J-P,
    2002 \textit{A\&A} \textbf{396}, L31

\item [Abrams, L. S.], 1979 \textit{Phys. Rev. D} \textbf{20}, 2474

\item [Abrams, L. S.], 1989 \textit{Can J. Phys.} \textbf{67}, 919, gr-qc/0102051

\item [Anderson, H. L.], 1989 in `A Physicists Desk Reference', Ed. AIP, p68

\item [Arons, J. et al.] 1984 in `High Energy Transients in Astrophysics',
    AIP Conf. Proc. 115, 215, Ed. S. Woosley, Santa Cruz, CA

\item [Barret, D. \& Olive, J-F.], 2002 \textit{ApJ} \textbf{576}, 391

\item [Belloni, T., Mendez, M., King, A.,
    van der Klis, M. \& van Paradijs, J.] 1997 \textit{ApJ} \textbf{488}, L109

\item [Baumgarte, T. \& Shapiro, S.] 2003 \textit{ApJ} \textbf{585}, 930

\item [Bhattacharya D.], \& Srinivasan, G., 1995 in `X-Ray Binaries',
    eds W. Lewin, J. van Paradijs \& E. van den Heuvel, Cambridge Univ. Press

\item [Bisnovatyi-Kogan, G.] \& Lovelace, R. 2000 \textit{APJ} \textbf{529}, 978

\item [Boyd, P. et al.] 2000 \textit{ApJ} \textbf{542}, L127

\item [Brazier, K. et al.], 1990 \textit{ApJ} \textbf{350}, 745

\item [Burderi, L., et al.] 2002 \textit{ApJ} \textbf{574}, 930

\item [Campana, S. et al.], 2002 \textit{ApJ} \textbf{580}, 389

\item [Campana, S. et al.], 1998 \textit{A\&A Rev.} \textbf{8}, 279

\item [Chakrabarty, D. et al.], 2003 \textit{Nature} \textbf{424}, 42

\item [Chou \& Tajima] 1999 \textit{ApJ} \textbf{513}, 401

\item [Church, M.J.] 2001 in Advances in Space Research, 33rd Cospar
    Scientific Assembly, Warsaw, Poland July 2000 astro-ph/0012411

\item [Church, M.] \& Balucinska-Church, M.
    2001 \textit{A\&A} \textbf{369}, 915

\item [Coburn, W.] \& Boggs, S., 2003 \textit{Nature} \textbf{423}, 415

\item [Corbel, S. ]\& Fender, R. 2002 \textit{ApJ} \textbf{573}, L35

\item [Cui, W.] 1997 \textit{ApJ} \textbf{482}, L163

\item [Done, C. \& $\dot{Z}$ycki, P.] 1999 \textit{MNRAS} \textbf{305}, 457

\item [Gallo, E., Fender, R., Pooley, G.], 2003 \textit{MNRAS}, \textbf{344}, 60

\item [Galloway, D., Chakrabarty, D., Morgan, E., \& Remillard, R.]
    2002 \textit{ApJL} \textbf{576}, 137

\item [Garcia, M. et al].. 2001 \textit{ApJ} \textbf{553}, L47

\item [Ghosh, P. \& Lamb, F. ]1992 in X-Ray Binaries
    and Recycled Pulsars, Ed. E. van den Huevel and S. Rappaport, Kluwer

\item [Gliozzi, M., Bodo, G. \& Ghisellini, G. ]1999 \textit{MNRAS} \textbf{303}, L37

\item [Gnedin, Y. et al.], 2003 Astrophys. Sp. Sci accepted, astro-ph/0304158

\item [Goodson, A., Bohm, K. \& Winglee, R.] 1999 \textit{ApJ} \textbf{534}, 142

\item [Gupta, A., Mishra, A., Mishra, H., \& Prasanna, A.R.], 1998, `Classical
    and Quantum Gravity', 15, 3131

\item [Harding, A.], 2003, Invited talk at Pulsars, AXPs  and SGRs
    Observed with BeppoSAX and Other Observatories, Marsala, Sicily, Sept. 2002
    astro-ph/0304120.

\item [Hawley, J., Balbus, S. \& Winters, W.], 1999 \textit{ApJ} \textbf{518}, 394

\item [Hernandez Jr., W.C.] \& Misner, C.W.,
    1966 \textit{ApJ}. \textbf{143}, 452

\item [Iaria, R. et al.] 2001 \textit{ApJ} \textbf{547}, 412

\item [Ibrahim, A., Swank, J. \& Parke, W.], 2003 \textit{ApJ} \textbf{584}, L17

\item [Ilarianov, A. \& Sunyaev, R.] 1975 \textit{A\&A} \textbf{39}, 185

\item [Kato, Y., Hayashi, M., Matsumoto, R.] 2004 ApJ in press
    astro-ph/0308437

\item [Kippenhahn, R. \& Weigert, A.] 1990 `Stellar
    Structure and Evolution' Springer-Verlag, Berlin Heidelberg New York

\item [Kluzniak, W. \& Ruderman, M], 1998 \textit{ApJL} \textbf{505}, 113

\item[Landau, L.D. \& Lifshitz, E. M.], 1975
    `Classical Theory of Fields', 4th Ed, Pergamon Press p248-252

\item [Landau, L. \& Lifshitz, E.] 1958
    `Statistical Physics', Pergamon Press LTD, London

\item [Leiter, D. \& Robertson, S.], 2003 \textit{Found Phys. Lett.} \textbf{16} 143

\item [Lindquist, R. W., Schwartz, R. A. \&
    Misner], C. W. 1965 \textit{Phys. Rev.}, \textbf{137}B, 1364.

\item [Lindquist, R. W.], 1966 \textit{Annals of Physics}, \textbf{37}, 487

\item [Livio, M., Ogilvie, G. \& Pringle, J. ]1999 \textit{ApJ} \textbf{512}, 100

\item [Maccarone, T., Gallo, E., Fender, R.] 2003 \textit{MNRAS} \textbf{345}, L19

\item [Markoff, S., Falcke H.,
    \& Fender, R.] 2001 \textit{A\&AL} \textbf{372}, 25

\item [Markoff, S., Nowak, M., Corbel, S., Fender, R., Falcke, H.], 2003 \textit{New Astron. Rev.} \textbf{47}, 491

\item [Markwardt, C.], et al. 2002 \textit{ApJL} \textbf{575}, 21

\item [Matt, S., Goodson, A., Winglee, R.], B\"{o}hm, K. 2002 \textit{ApJ} \textbf{574}, 232

\item [Mauche, C. W.], 2002 \textit{ApJ} \textbf{580},423

\item [McClintock, J. et al.], 2001 \textit{ApJ} \textbf{556}, 42

\item [Merloni, A., Fabian, A.], 2002 \textit{MNRAS}, \textbf{332}, 165

\item [Merloni, A., Heinz, S., Di Matteo, T.], 2003 \textit{MNRAS}, \textbf{345}, 1057

\item [Miller, J. et al.], 2003 \textit{ApJL}, \textbf{583}, 99

\item [Misner, C. W. ]1965 \textit{Phys. Rev.}, \textbf{137}B, 1360

\item [Misner, C.], Thorne, K. \& Wheeler, J. 1973 `Gravitation', Freeman, San Francisco, California

\item [Mitra, A.] 1998 \textit{ApJ} \textbf{499}, 385

\item [Mitra, A.] 2000 \textit{Found. Phys. Lett.} \textbf{13}, 543

\item [Mitra, A.] 2002 \textit{Found. Phys. Lett.} \textbf{15},
439

\item[Mitra, A.] 2005, Talk given at the 29th International Cosmic
Ray Conference in Pune India, Astro-ph/0506183

\item[Mitra, A.] 2005 Astro-ph/0512006

\item[Mitra, A.] in press 2006 in `Focus On Black Hole Research',
ed. P.V. Kreitler, Nova Science Publishers, Inc. ISBN
1-59454-460-3, novapublishers.com)

\item [Mitra, A.] 2006 \textit{MNRAS Letters} in press,
gr-qc/0601025

\item [Muno, M., Remillard, R. \& Chakrabarty, D.]
    2002 \textit{ApJ} \textbf{568}, L35

\item [Narayan, R., Garcia, M., McClintock, J.], 1997 \textit{ApJ}. \textbf{478}, L79

\item [Orosz, J., Remillard, R., Bailyn, C.] \& McClintock, J.
    1997 \textit{ApJ} \textbf{478}, L83

\item [Paragi, Z., et al.] 2002 Talk presented
    at the 4th Microquasar Workshop, Cargese, Corsica May 27-31  astro-ph/0208125

\item [Pelletier, G. \& Marcowith, A.], 1998 \textit{ApJ} \textbf{502}, 598

\item [Possenti, A., Cerutti, R., Colpi, M.,
    \& Mereghetti, S.] 2002 \textit{A\&A} \textbf{387}, 993

\item [Punsly, B.] 1998 \textit{ApJ} \textbf{498}, 640

\item [Psaltis, D., Belloni, T.], \& van der Klis, M., 1999 \textit{ApJ} \textbf{520},262

\item [Reid, M. et al.], 2003 \textit{ApJ} \textbf{587}, 208

\item [Robertson, S. \& Leiter, D.] 2002 \textit{ApJ}, \textbf{565}, 447

\item [Robertson, S., Leiter, D.] 2003 \textit{ApJ}, \textbf{596}, L203

\item [Robertson, S., Leiter, D.] 2004 \textit{MNRAS} \textbf{350}, 1391

\item[Robertson, S., and Leiter, D.] 2005 `The Magnetospheric
Eternally Collapsing Object (MECO) Model of Galactic Black Hole
Candidates and Active Galactic Nuclei', pp 1-45 (in New
Developments in Black Hole Research, ed. P.V.Kreitler, Nova
Science Publishers, Inc. ISBN 1-59454-460-3, novapublishers.com)

\item [Schild, R., Leiter, D. \& Robertson] submitted to
\textit{AJ} astro-ph/0505518

\item [Shapiro, S. \& Teukolsky, S.] 1983 in
    `Black Holes, White Dwarfs \& Neutron Stars', John Wiley \& Sons, Inc., New York

\item [Soria, R., Page, M. \& Wu, K.] 2002 in Proceedings of the Symposium
    `New Visions of the X-Ray Universe in the XMM-Newton \& Chandra Era',
    Nov. 2001, ESTEC, the Netherlands, astro-ph/0202015

\item [Stella, L., White, N. \& Rosner, R.] 1986 \textit{ApJ} \textbf{308}, 669

\item [Strohmayer, T. \& Markwardt, C.], 2002 \textit{ApJ} \textbf{577}, 337

\item [Sunyaev, R.], 1990 Soviet Astron. Lett. 16, 59

\item [Tanaka, Y.\& Lewin, W.], 1995 in `Black-hole binaries'
    in X-ray Binaries ed. W. Lewin, J. van Paradijs\& E. van den Heuvel
    (Cambridge: Cambridge Univ. Press)

\item [Tanaka, Y. \& Shibazaki, N.], 1996 \textit{ARA\&A}, \textbf{34}, 607

\item [Thorne, K.], 1965 \textit{Phys. Rev}, \textbf{138}, B251

\item [Titarchuk, L.] \& Wood, K., 2002 \textit{ApJL} \textbf{577}, 23

\item [Uzdensky, D.], 2002 \textit{ApJ} \textbf{572}, 432

\item [Vadawale, S., Rao, A. ]\& Chakrabarti, S. 2001 \textit{A\&A} \textbf{372}, 793V

\item [van der Klis, M.] 1994 \textit{ApJS} \textbf{92}, 511

\item [Warner, B. \& Woudt, P.], 2003 in ASP Conf. Series, Ed. M. Cuppi
    and S. Vrielmann astro-ph/0301168


\item [Wheeler, J. \& Ciufolini, I.], 1995 in `Gravitation And Inertia'
    Princeton Univ. Press, 41 William St., Princeton, New Jersey

\item [White, N.\& Marshall, F.] 1984, \textit{ApJ}. \textbf{281}, 354

\item [Wilson, C. \& Done, C.] 2001 \textit{MNRAS} \textbf{325}, 167

\item[Zaumen, W. T. ], 1976 \textit{ApJ}, \textbf{210}, 776

\item [Zhang, W., Yu, W. \& Zhang, S.] 1998 \textit{ApJ} \textbf{494}, L71

\item [$\dot{Z}$ycki, P., Done, C. and Smith], D. 1997a in AIP
    Conf. Proc. 431, Accretion Processes in Astrophysical Systems:
    Some Like It Hot, Ed. S. S. Holt \& T. R. Kallman (New York, AIP), 319

\item $\dot{Z}$ycki, P., Done, C. and Smith 1997b \textit{ApJ} \textbf{488}, L113
\end{refhanging}


\end{document}